\documentclass[preprint]{aastex}
\usepackage{ulem}
\usepackage{graphics}
\usepackage{epstopdf}

\usepackage{url}
\usepackage{caption2}
\usepackage{lscape}
\usepackage{enumerate}
\usepackage{color}

\shorttitle{Observing Conditions}

\shortauthors{Tian et al.}

\begin{document}

%\title{50BiN open cluster survey in time-domain: site condition and infrastructure}
\title{Optical Observing Conditions at Delingha Station}

\author{J.F. Tian\altaffilmark{1,2}, L.C. Deng\altaffilmark{1}, X.B. Zhang\altaffilmark{1}, X.M. Lu\altaffilmark{1}, J.J. Sun\altaffilmark{4}, Q.L. Liu\altaffilmark{4}, Q. Zhou\altaffilmark{4}, Z.Z. Yan\altaffilmark{1,2},
	Y. Xin\altaffilmark{1}, K. Wang\altaffilmark{1,2},  X.J. Jiang\altaffilmark{1},
	Z.Q. Luo\altaffilmark{3}, J. Yang\altaffilmark{4}}

\affil {1, Key Laboratory of Optical Astronomy, National Astronomical Observatories, Chinese Academy of Sciences, Beijing, 100012, China}
\affil {2, University of Chinese Academy of Sciences, Beijing 100049, China}
\affil {3, Physics and Space Science College, China West Normal University, Nanchong 637002, China}
\affil {4, Purple Mountain Observatory, Chinese Academy of Sciences, Nanjing, 210008, China}

\begin{abstract}

SONG is a global ground based network of 1 meter telescopes for stellar time-domain science, an international collaboration involving many countries across the world. In order to enable a favourable duty cycle, the SONG network plans to create a homogeneous distribution of 4 nodes in each of the northern and southern hemispheres. A natural possibility was building one of the northern nodes in East Asia, preferably on the Qinghai-Tibetan Plateau. During the last decade, a great deal of effort has been invested in searching for high a quality site for ground based astronomy in China, since this has been one of the major concerns for the development of Chinese astronomy.  A number of sites on the plateau have been in operation for many years, but most of them are used only for radio astronomy, as well as small optical telescopes for applied astronomy. Several potential sites for large optical instruments have been identified by the plateau site survey, but as yet none of them have been adequately quantitatively characterised. Here we present results from a detailed multi-year study of the Delingha site, which was eventually selected for the SONG-China node.  We also describe the site monitoring system that will allow an isolated SONG and 50BiN node to operate safely in an automated mode.

\end{abstract}

%\keywords{Data Analysis and Techniques}

%\section{Introduction} \label{intro}
\section{INTRODUCTION} \label{intro}
%SONG/50BiN global site plan. site survey in west china, identify the site.

Asteroseismology via high resolution spectroscopy is a more powerful tool compared with astrometry via photometry, since spectroscopy enables direct measurement of the surface velocities. However, spectroscopic asteroseismology from the ground requires a network of instruments. The Stellar Observations Network Group (SONG) is an ambitious project that consists of 8 telescope nodes which aims to study edge-cutting problems, namely determining detailed internal structures of oscillating stars and looking for Earth-like planetary systems revealed by micro-lensing events.  In order to achieve a high duty cycle within a limited budget and to produce homogenous data, a network of identical 1 meter telescope and instruments is planned, which includes 4 nodes almost homogeneously distributed in geographic latitude on each hemisphere \citep{2008IAUS..252..465G}. To form such a network in the north, one node must be located in East Asia. Then it is straightforward to plan the remaining 3 sites, in terms of geographic distribution and site quality, in the Canary Islands, the Hawaiian islands and the western Pacific coast of the US.  However, due to the global climate pattern, it is unclear whether East Asia is likely to contain a site that can match the quality of those other three oustanding locations, which is necessary to optimize the network's duty-cycle.  The Qinghai-Tibetan Plateau provides our best chance to find a site that can deliver data of both quality and quantity that matches the other nodes.

During the last decade, the search for a high quality site for ground based astronomy has been one of the major concerns for the development of general-purpose astronomy in China.  A great deal of effort has been invested towards this goal over recent years.  Since the Qinghai-Tibetan Plateau has a 
high altitude, dry climate and obviously low light pollution, it has naturally been searched for future high-quality sites. A specific research group to perform site surveys in the west land of western China has been organized \citep{2005JKAS...38..113Y}. Some sites, including Oma and Ali, were then selected for detailed site testing measurements \citep{2012SPIE.8444E..1KY}.

This paper is to present the results of a qualification campaign for the Delingha site, one of the pre-selected potential locations for the SONG project, which is organized as follows. \S\ \ref{song-site} gives the basic requirements of a SONG node, with specific limitations for the node in East-Asia. \S\ \ref{about}  describes the general nature of the site, including the location, infrastructure, climate and environment. \S\  \ref{condition} presents the data from the site survey regarding the optical observing conditions. \S\ \ref{infra} describes how the environmental monitoring system works.  \S\ \ref{conclusion} presents the summary and conclusions.

\section{A SONG NODE in CHINA}\label{song-site}

China joined SONG in 2009 with a promise to find a suitable site for a SONG node and to build up a full set of instruments.  China spans a huge range in longitude, and for identifying a suitable site for a SONG node the most important issues are the general weather pattern (in terms of good observing nights per year) and the existence of supporting infrastructure (e.g., accessibility, power supply, Internet), since the project budget is limited.

Consideration of the weather patterns and basic observing condition criteria suggests that we should investigate the western part of the country, on the Qinghai-Tibetan plateau.  There are a number of existing astronomical observing sites in this region, mostly operating radio telescopes.  The characteristics of the pre-selected candidate sites are given in Table \ref{tbl-sites}. There are also a few new sites under testing in the region. Of course it would be preferable to use one of the existing sites, because no significant infrastructure construction work would be needed.  More importantly, existing technical support would then be available for emergency and routine maintenance \citep[as suggested by][]{2008IAUS..252..465G}.  Although the western China sites in Table \ref{tbl-sites} have been in operation for many years, most of them are used only for radio astronomy, as well as for small optical telescopes for applied astronomy purposes (e.g., for tracking space debris). Their feasibility for high quality optical-infrared astronomical observations had never been systematically tested, and so their suitability needed to be investigated before we could determine a site for the Chinese node of SONG.

A general-purpose site survey has been running in the area for many years, and a number of new sites have been identified. However, the SONG timeline and requirements preclude all new immature sites, such as those in the Oma and Ali areas. This is even though, e.g., there are indications showing that Ali may well be a better site for IR-optical astronomy than the other known sites \citep{2012SPIE.8444E..1KY}.  But all of those locations are still being tested, and unfortunately have no supporting infrastructure that could fulfill our requirements.  On the other hand, we cannot use the well-established sites in the more developed regions of China. These are mostly in heavily populated regions of the country near the eastern coast, and all of them are limited by observing conditions (especially the number of clear nights, due to climate patterns) and subject to severe light pollution.  This includes the Xinglong site where LAMOST is located. 

\begin{table}
	\caption{A list of the preselected sites for SONG in West China}\label{tbl-sites}
	\center
	\begin{tabular}{lcccl}
		\hline
		Site & Longitude & Latitude & Elevation & Description \\ \hline
		Nanshan (Xinjiang) & $81^{\circ} 10\arcmin.5$E  & $43^{\circ} 28\arcmin.4$N & 2080m & Radio \& Optical \\
		Ali (Tibet) & $80^{\circ} 01\arcmin.2$E  & $32^{\circ} 19\arcmin.2$N  & 5100m & Optical \& IR (not ready) \\
		Yangbajing (Tibet) & $90^{\circ} 31\arcmin.5$E & $30^{\circ} 06\arcmin.2$N   & 4300m & Cosmic ray \& sub-mm\\
		Gaomeigu (Yunnan) & $100^{\circ} 01\arcmin.9$E & $26^{\circ} 42\arcmin.5$N & 3200m & Optical\\
		Delingha (Qinghai) & $97^{\circ} 33\arcmin.6$E & $37^{\circ} 22\arcmin.4$N & 3200m & Radio \& Optical\\ \hline
	\end{tabular}
\end{table}

A good site for a node of SONG should comply with the basic requirements regarding the network's duty cycle and data quality.  In the northern hemisphere, the other three nodes are located in world-class sites. Their outstanding quality makes it challenging to find a site which is comparable to them, as is our aim. In principle, a good site for a SONG project telescope should have long-term statistics demonstrating a high percentage of clear nights, dry air, stable atmospheric conditions (good seeing) and low light pollution (dark night sky), as described in \citet{2013IAUS..288..318D}.  For the specific site of the Chinese SONG node, the order of the selection criteria is as follows: the number of clear/partial clear nights per year, the site infrastructure and technical support resources, the seeing, and the night sky brightness.  The meteorological variables to be studied for site characterization are: air temperature, relative humidity and wind velocity (including speed and direction). The critical site parameters for astronomical observations that need to be studied are the number of clear nights, the night sky brightness and the seeing.

This article summarizes the site qualification campaigns over the past five years, in which we present the conditions for optical observations from this site. In addition to SONG's original plan \citep{2008IAUS..252..465G}, the Chinese team proposed a photometry subnetwork named the ``50 cm Binocular Network'' (50BiN).  50BiN will co-locate telescopes with the 4 nodes of SONG in the northern hemisphere to enable a large field photometric capability for the network  \citep{2013IAUS..288..318D,2014AJ....148...40Z,2014AJ....148..106Z,2015AJ....150..161W}. For the SONG project, the 1m telescope and the spectrograph of the Chinese SONG node are fabricated at NIAOT \citep{2012SPIE.8444E..5TW} based on an identical design to the other nodes.  Unlike the other SONG sites in the northern hemisphere, the Chinese SONG instruments are located at a radio site, and in an isolated environment.  In order to help the isolated SONG and 50BiN node to run in an automated mode, it is necessary to install instruments for detecting the state of the environment.  The environmental instruments now installed in Delingha were produced by various companies, although it was necessary to improve both the software and the hardware of these instruments to produce an independent system which meets our requirements. In this article we also describe how that monitoring system works.

%\section{Site description}  \label{about}
\section{SITE DESCRIPTION}  \label{about}

Delingha observing station\footnote{\url{http://english.dlh.pmo.cas.cn/}} lies about 40 km to the east of the city of Delingha; we will refer to it as the Delingha site. It is operated by the Purple Mountain Observatory as a site for radio astronomy. The Delingha site was established in the early 1980s, and has long had a power supply from the grid, along with other supporting facilities.  The site is located on the Qinghai-Tibetan Plateau, with specific position as given in Table \ref{tbl-sites}. Its primary facility is a 13.7 m radio telescope working in the millimeter band, and it also hosts a few small optical telescopes used for observations of celestial objects in the solar system. The site infrastructure is adequate to meet our requirements.  A special optical fiber network connection has been available since 2001; this facilitates good communication with the outside world, including telephone connections and video conferencing. There a guest house and a canteen to support staff and visitors. A team of technicians is always available for both emergency situations and scheduled maintenance. The site is connected to the city of Delingha by a highway.  Delingha city, which is the third largest city of Qinghai Province, has multiple good transport links. In terms of connectivity and infrastructure, this site is perfect for a SONG node.

Delingha has a plateau continental climate with an oxygen deficit, cold and dry air, and receives very little rain. In the early 1980s when the site was selected for radio observations \citep{1985AcASn..26..187H}, the local meteorological conditions were described as follows: ``The annual average temperature is $-0.7\,^{\circ}\mathrm{C}$. In winter the average temperature is $-15.1\,^{\circ}\mathrm{C}$ and in summer the average temperature is $13.0\,^{\circ}\mathrm{C}$. The extreme low temperature recorded is $-38.9\,^{\circ}\mathrm{C}$. The annual average relative humidity is $43\%$. In winter the average relative humidity is $32.9\%$ and in summer the average relative humidity is $45.8\%$. The number of fully clear days and nights is 108 days and totally more than 300 days of partially clear sky.  The average wind speed is $2.1~{\rm m~s^{-1}}$ and the prevailing wind direction at the surface level is south-west. The annual average precipitation is $158~\rm mm$, mainly in summer.'' These data indicate that Delingha is potentially a good site for optical observations, and in 2010 the community reached a consensus to monitor the site in detail.

%\section{Natural observing conditions}  \label{condition}
\section{NATURAL CONDITIONS FOR ASTRONOMICAL OBSERVATIONS}  \label{condition}

The fundamental ground-level meteorological properties for site characterization are the air temperature, barometric pressure, relative humidity and wind speed and direction \citep{2000A&AS..147..271J}. In this paper we will focus on long term statistics of air temperature, relative humidity and wind speed and direction. Air temperature is important for the operation of telescopes, detectors and other instruments. The minimum operating temperature even for industrial grade instruments is typically $-40\,^{\circ}\mathrm{C}$. Relative humidity is also important for determining whether or not astronomical observations can be made.  When the humidity is higher than $90\%$ observations should stop \citep{1985VA.....28..449M}, since the optical surfaces can become wet and electronic equipment can be damaged. Hence the percentage of hours with humidity less than $90\%$ sets an upper limit on observing time \citep{2000A&AS..147..271J}. Wind speed and direction are also crucial when operating telescopes.  \citet{1985VA.....28..449M} suggested the maximum safe operating value is $15~{\rm m~s^{-1}}$, in which case the percentage of hours with wind speed less than $15~{\rm m~s^{-1}}$ sets another upper limit on usable observing time.

For a site dedicated to a ground based network for time-domain tasks whose duty-cycle and data quality are critical, the most important site parameters are the number of clear nights, the seeing, and the sky brightness. As far as a selecting a SONG site in East Asia is concerned, the total number of usable observing nights is the top requirement.  Whether a night is suitably clear or not can be judged directly from the cloud coverage. We use an all sky camera (ASC hereafter) to monitor the clould coverage of the sky during the day as well as in the night. \citet{2014arXiv1402.4762M} and \citet{2013IAUS..288...38S}  also used an ASC for night sky monitoring.  Night sky brightness restricts the signal-to-noise ratio that the telescope can reach for a given exposure time. The major contributions to the night sky brightness are from light pollution, moonlight and other natural backgrounds. We use a Sky Quality Meter (SQM), a widely used commercial device, to monitor the night sky brightness. The same device has been used for the campaign of ``GLOBE at Night'' \citep{2008ASPC..400..152S}. Seeing plays an import role in optical observing, and degradations to seeing are mainly caused by  atmospheric turbulence.  Seeing can be measured in many different ways. We adopted a rather simple approach, DIMM (Differential Image Motion Monitor), which was introduced by \citet{1990A&A...227..294S}.  In this case, seeing is quantitatively measured via the differential positions of two sub-aperture images of a star.  At the beginning of site testing, a SBIG  seeing monitor (a nonprofessional instrument) was also used to provide a fast preview.

%\subsection{climate and weather pattern}
\subsection{Climate and Weather Pattern}

%\textbf{\emph{Stephen:  I think ``Hpa'' should probably be changed to ``hPa'', but I haven't done so in case I'm wrong. I also suspect that PASP would prefer one of $m~s^{-1}$ or $\rm m~s^{-1}$ to $m/s$, but I haven't checked the PASP preferences. Also, doesn't throwing out 3$\sigma$ changes in wind speed reduce your information about gusty winds (especially if you are using the measurement accuracies as $\sigma$ for the new weather station, in which case $3\sigma < 1~{\rm m~s^{-1}}$)?}}

A 4-element laboratory-made meteorological station  was deployed at Delingha site in the period from  September 2010 to June 2013.  Installed on a top of a 10 meter high tower, it measured and recorded air temperature, relative humidity,  barometric pressure and wind speed. The weather data it collected was archived. The sensors are all calibrated to the following specified absolute measurement accuracies: air temperature $ \pm0.5\,^{\circ}\mathrm{C}$, relative humidity $\pm1.8\% $, pressure  $\pm3.2~\rm hPa$ , wind speed  $\pm3.2~{\rm m~s^{-1}}$.  The weather station recorded data once per minute.   A commercial  high precision ultrasonic meteorological station\footnote{\url{http://www.jz322.net/prdouct_text.aspx?id=297}} was used to replace the old one in June 2013.  The new weather station can measure and record temperature, relative humidity, barometric pressure, dew point temperature, wind speed and wind direction. The manufacturer's specified absolute measurement accuracies are: air temperature $ \pm0.1\,^{\circ}\mathrm{C}$, relative humidity $\pm2\% $, pressure  $\pm0.3~\rm hPa$, dew point temperature $ \pm0.2\,^{\circ}\mathrm{C}$,  wind speed  $\pm0.3~{\rm m~s^{-1}}$,  wind direction  $\pm3^{\circ}$.  We have analyzed all the data from the two weather stations, which is summarized below.  Abnormal data points deviating from adjacent data by more than 3$\sigma$ were filtered out.

\subsubsection{Air Temperature}

% In this paper, we defined the time after beginning of astronomical morning twilight and before end of astronomical evening twilight as daytime,

\begin{figure}
		\begin{center}
			\includegraphics[angle=0, width=1\textwidth]{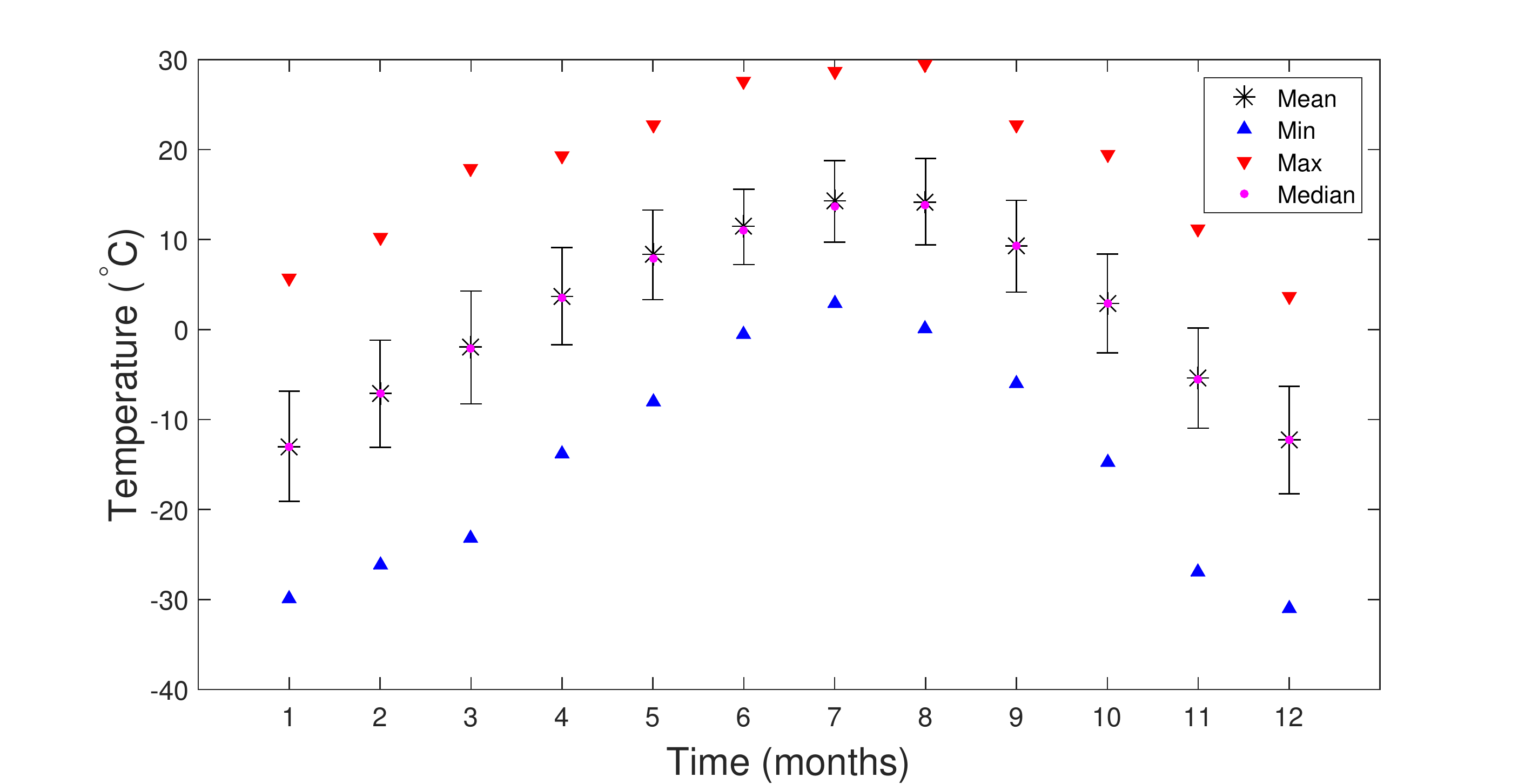}
			\caption{Monthly statistics of the air temperature from 2010 to 2014. See the electronic edition of the PASP for a color version of this figure.}
			\label{fig-month-temperature}
		\end{center}
	\end{figure}

\begin{figure}
		\begin{center}
			\includegraphics[angle=0, width=1\textwidth]{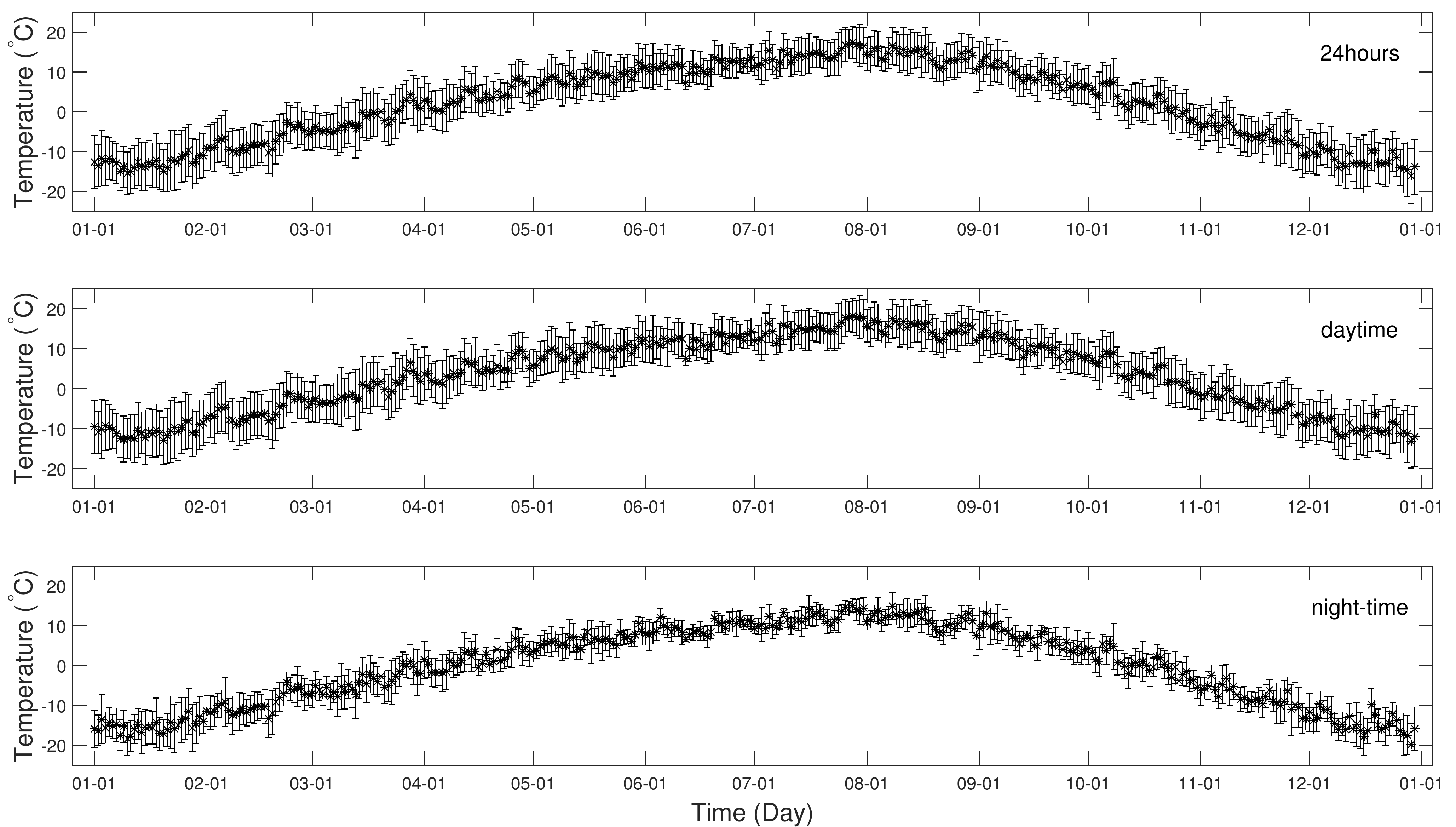}
%			\caption{Daily mean of air temperature during all day and its standard deviation from 2010 to 2014}
			\caption{Mean air temperatures and their standard deviations from 2010 to 2014, averaged over 24 h, daytime and night-time periods (in the top, middle and bottom panels, respectively). }
			\label{fig-temperature_all}
		\end{center}
	\end{figure}

We have analyzed the distribution of air temperature by month and calculated the maximum, minimum, mean, median and standard deviation values that are shown in Figure \ref{fig-month-temperature}. The average temperature is $1.8\,^{\circ}\mathrm{C}$ with a maximum of $29.4\,^{\circ}\mathrm{C}$ and minimum of $-31.0\,^{\circ}\mathrm{C}$. The absolute difference between the maximum and minimum temperature during each month are broadly the same, and the mean and median values of the air temperature during each month are very similar.  We also analyze the difference in the air temperature between the day and night.  For this purpose we define the time from the beginning of astronomical morning twilight to the end of astronomical evening twilight as daytime, and defined the time from the end of astronomical evening twilight of this day to the beginning of astronomical morning twilight of the next day as the night-time of this day. Monthly statistics on daytime and night-time data of air temperature are  summarized in Table \ref{tbl-month-temperature}. Mean and standard deviation air temperatures for all 24 h, daytime and night-time periods from 2010 to 2014 are shown in Figure \ref{fig-temperature_all}.

%	\begin{figure}
%		\begin{center}
%			\includegraphics[angle=0, width=1\textwidth]{f3.eps}
%			\caption{Daily mean of air temperature during daytime and its standard deviation for the whole year from 2010 to 2014}
%			\label{fig-temperature_day}
%		\end{center}
%	\end{figure}
%	
%	
%	
%	\begin{figure}
%		\begin{center}
%			\includegraphics[angle=0, width=1\textwidth]{f4.eps}
%			\caption{Daily mean of air temperature during night-time and its standard deviation from 2010 to 2014}
%			\label{fig-temperature_night}
%		\end{center}
%	\end{figure}

	\begin{deluxetable}{lcccccccccccc}
		\tablecolumns{13}
%		\tabletypesize{\scriptsize}
%		\rotate
%		\tablewidth{50pc}
		\tablecaption{Monthly statistics on daytime and night-time data of air temperature during 2010-2014 \label{tbl-month-temperature}}
		\tablehead
		{
				\colhead{}  &   \colhead{} &  \multicolumn{5}{c}{Daytime Temperature $\left(\,^{\circ}\mathrm{C} \right)$ }  &   \colhead{}  & \multicolumn{5}{c}{Night-time Temperature $\left(\,^{\circ}\mathrm{C} \right)$}  \\
				 \cline{3-7} \cline{9-13} \\
				\colhead{Month} & \colhead{} & \colhead{Max}   & \colhead{Min}    & \colhead{Median} & \colhead{Mean}   & \colhead{Std} & \colhead{} & \colhead{Max}   & \colhead{Min}& \colhead{Median}  & \colhead{Mean}   & \colhead{Std}
		 }
		\startdata
			 Jan. & & 5.7 & -29.9 & -10.2 & -10.8 & 6.3 & & 1.4 & -29.1 & -15.8 & -15.4 & 4.8\\
			 Feb. & & 10.3 & -26.2 & -4.7 & -5.3 & 6.2 & & 4.1 & -25.8 & -9.2 & -9.4 & 4.6\\
			 Mar. & & 17.9 & -23.2 & -0.1 & -0.4 & 6.5 & & 12.3 & -21.7 & -4.5 & -4.3 & 4.8\\
			 Apr. & & 19.3 & -13.9 & 5.1 & 4.9 & 5.5 & & 13.9 & -12.2 & 1.2 & 1.4 & 4.1\\
			 May. & & 22.7 & -8.0 & 9.1 & 9.1 & 5.2 & & 16.3 & -5.7 & 6.0 & 6.1 & 3.5\\
			 Jun. & & 27.6 & -0.5 & 11.9 & 12.0 & 4.3 & & 16.9 & 1.8 & 9.5 & 9.4 & 2.7\\
			 Jul. & & 28.7 & 2.9 & 14.5 & 15.0 & 4.7 & & 22.3 & 4.6 & 11.9 & 12.0 & 3.0\\
			 Aug. & & 29.4 & 0.1 & 15.1 & 15.2 & 4.9 & & 22.2 & 0.5 & 11.6 & 11.7 & 3.4\\
			 Sep. & & 22.8 & -6.0 & 11.2 & 10.6 & 5.1 & & 19.7 & -5.6 & 6.5 & 6.8 & 4.1\\
			 Oct. & & 19.5 & -14.8 & 5.4 & 4.4 & 5.7 & & 15.1 & -13.2 & 0.6 & 0.6 & 4.3\\
			 Nov. & & 11.2 & -24.7 & -2.8 & -3.5 & 5.8 & & 4.8 & -27.0 & -7.6 & -7.8 & 4.5\\
			 Dec. & & 3.6 &  -31 & -8.9 & -10.0 & 6.1 & & -0.3 & -29.4 & -14.6 & -14.8 & 4.8\\
			 Total & & 29.4 & -31.0 & 2.8 & 4.3 & 10.4 & & 22.3 & -29.4 & -2.6 & -2.4 & 10.3\\
%			 \cutinhead{This is a cut-in head}

		\enddata

%		\tablenotetext{a}{with the deluxetable environment}
%		\tablecomments{Tabl}
		
	\end{deluxetable}

\subsubsection{Relative Humidity}

\begin{figure}
		\begin{center}
			\includegraphics[angle=0, width=1\textwidth]{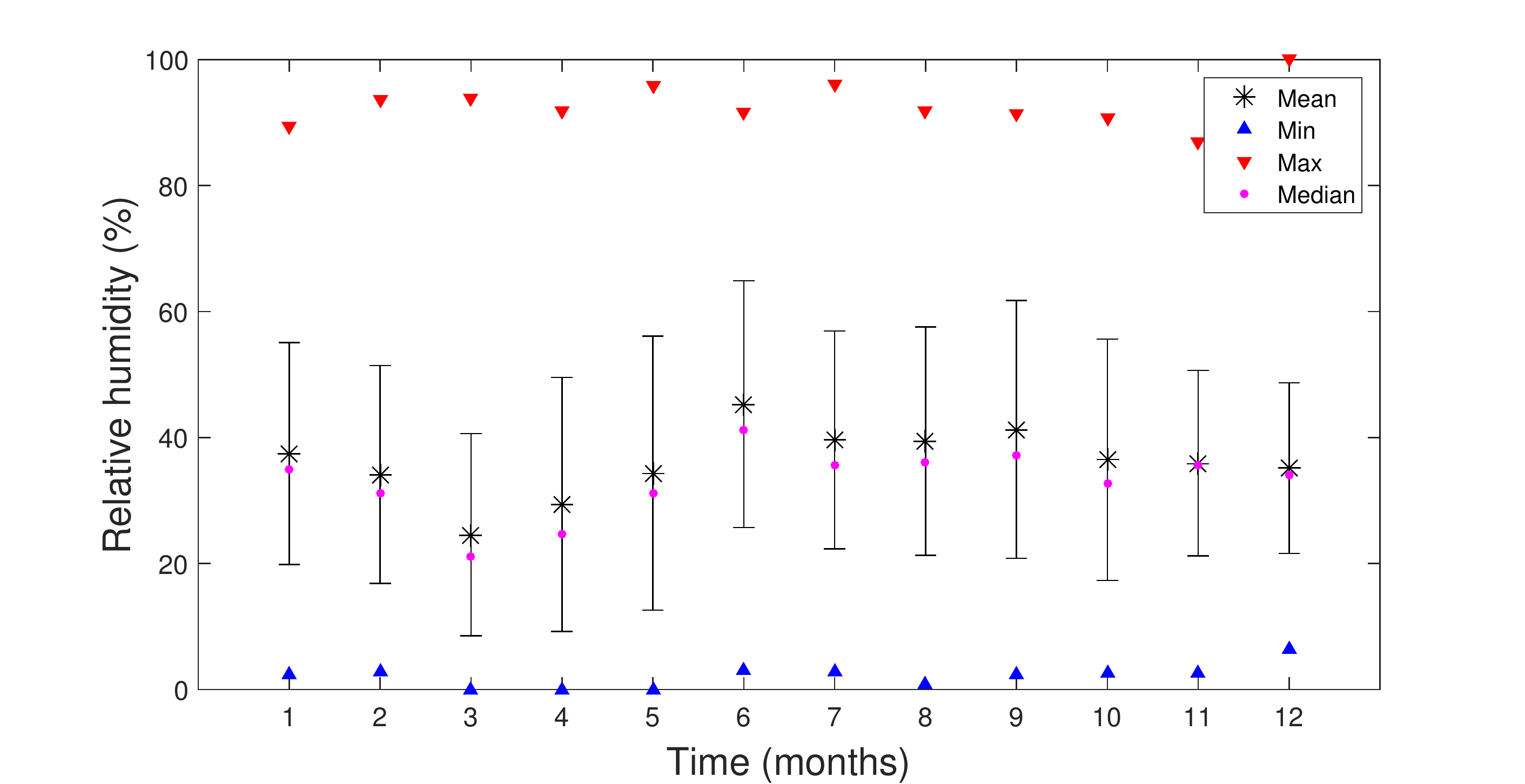}
			\caption{Monthly statistics of the relative humidity  from 2010 to 2014. See the electronic edition of the PASP for a color version of this figure.}
			\label{fig-month-humidity}
		\end{center}
	\end{figure}

\begin{figure}
		\begin{center}
			\includegraphics[angle=0, width=1\textwidth]{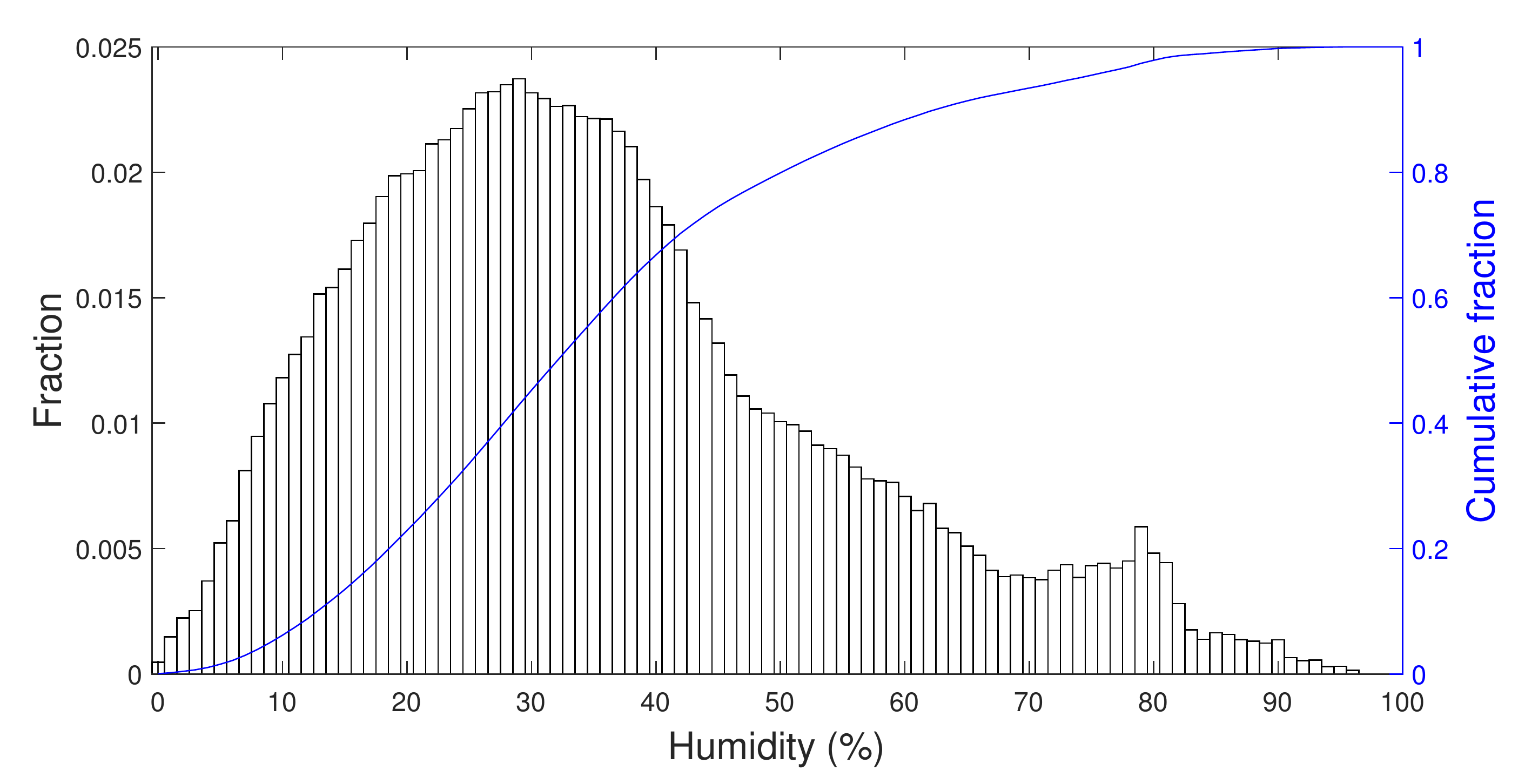}
			\caption{Humidity histograms and cumulative distributions from 2010 to 2014.  See the electronic edition of the PASP for a color version of this figure.}
			\label{fig-humidity-t}
		\end{center}
	\end{figure}

\begin{figure}
		\begin{center}
			\includegraphics[angle=0, width=1\textwidth]{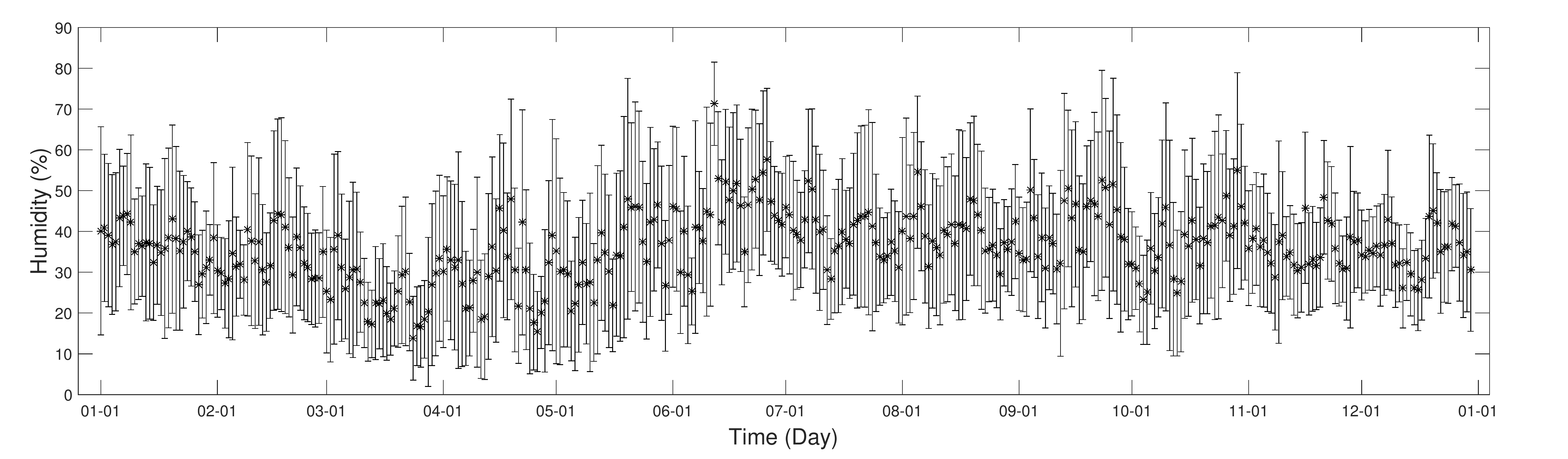}
			\caption{24 h mean relative humidity and standard deviations from 2010 to 2014}
			\label{fig-data-humidity}
		\end{center}
	\end{figure}

The site was selected for a millimeter-band radio telescope, for which it is desirable that the water column density should be low.  However, it is obvious that the humidity changes both from season-to-season and over longer time-scales. We have analyzed the monthly distribution of the relative humidity and calculated the maximum, minimum, mean, median and standard deviation values that are shown in Figure \ref{fig-month-humidity}. The average value of the relative humidity is $33.1\%$ .  Sometimes the humidity is very high, even higher than $90\%$, which precludes observations \citep{1985VA.....28..449M}. Frost may form on metal surfaces at this site, especially during the winter, when the relative humidity reaches 60\% or higher at low temperatures; while in summer, moisture may accumulate on such surfaces. This situation usually occurs on cloudy, rainy or snowy days and may continue for a short period. However, the humidity is usually less than $50\%$, and the median value is $32.6\%$, as illustrated in Figure \ref{fig-humidity-t}. Monthly statistics on the daytime and night-time data regarding relative humidity are summarized in Table \ref{tbl-month-humidity}. The relative humidity during night-time is typically higher than that during daytime, but the night-time relative humidity still has a median value of only $37.1\%$.  As expected from the climate of the site, the relative humidity is higher in summer, which brings summer rain. The 24-hour mean relative humidity (and standard deviations) are shown in Figure \ref{fig-data-humidity}.

	\begin{deluxetable}{lcccccccccccc}
		\tablecolumns{13}
%		\tabletypesize{\scriptsize}
%		\rotate
%		\tablewidth{50pc}
		\tablecaption{Monthly statistics on daytime and night-time data of  relative humidity during 2010-2014 \label{tbl-month-humidity}}
		\tablehead
		{
				\colhead{}  &   \colhead{} &  \multicolumn{5}{c}{Daytime Humidity $\left(\% \right)$ }  &   \colhead{}  & \multicolumn{5}{c}{Night-time Humidity $\left(\% \right)$}  \\
				 \cline{3-7} \cline{9-13} \\
				\colhead{Month} & \colhead{} & \colhead{Max}   & \colhead{Min}    & \colhead{Median} & \colhead{Mean}   & \colhead{Std} & \colhead{} & \colhead{Max}   & \colhead{Min}& \colhead{Median}  & \colhead{Mean}   & \colhead{Std}
		 }
		\startdata
			 Jan. & & 89.5 & 2.3 & 30.8 & 33.7 & 17.5 & & 88.7 & 6.1 & 38.9 & 41.9 & 16.6\\
			 Feb. & & 93.3 & 2.7 & 27.9 & 31.1 & 17.3 & & 93.7 & 8.0 & 35.0 & 38.2 & 16.4\\
			 Mar. & & 93.8 & 0 & 18.5 & 22.4 & 16.1 & & 91.1 & 1.6 & 24.3 & 27.9 & 15.3\\
			 Apr. & & 91.8 & 0 & 22.1 & 27.0 & 19.8 & & 89.2 & 0 & 29.4 & 34.1 & 19.8\\
			 May. & & 95.7 & 0 & 28.7 & 32.5 & 21.7 & & 95.8 & 1.7 & 35.5 & 39.3 & 20.9\\
			 Jun. & & 90.2 & 3.1 & 39.1 & 43.8 & 19.9 & & 91.7 & 12.5 & 48.2 & 50.7 & 17.3\\
			 Jul. & & 96.2 & 2.8 & 33.2 & 37.7 & 17.3 & & 96.0 & 14.7 & 42.7 & 45.1 & 16.1\\
			 Aug. & & 91.9 & 0.7 & 32.7 & 37.0 & 18.2 & & 90.2 & 7.5 & 43.0 & 45.2 & 16.5\\
			 Sep. & & 91.2 & 2.4 & 33.4 & 37.9 & 20.2 & & 91.4 & 8.0 & 40.8 & 46.7 & 19.5\\
			 Oct. & & 90.7 & 2.5 & 27.7 & 33.2 & 19.6 & & 87.8 & 7.6 & 38.3 & 41.4 & 17.6\\
			 Nov. & & 86.9 & 2.6 & 29.1 & 32.7 & 15.9 & & 84.8 & 9.9 & 38.7 & 39.7 & 12.1\\
			 Dec. & & 100 &  6.3 & 29.7 & 31.3 & 13.4 & & 100 & 13.4 & 37.9 & 39.2 & 12.4\\
			 total & & 100 & 0 & 29.6 & 33.1 & 19.1 & & 100 & 0 & 37.1 & 39.7 & 17.6\\
%			 \cutinhead{This is a cut-in head}

		\enddata

%		\tablenotetext{a}{with the deluxetable environment}
%		\tablecomments{Tabl}
		
	\end{deluxetable}

\subsubsection{Wind Speed and Direction}	
\begin{figure}
		\begin{center}
			\includegraphics[angle=0, width=1\textwidth]{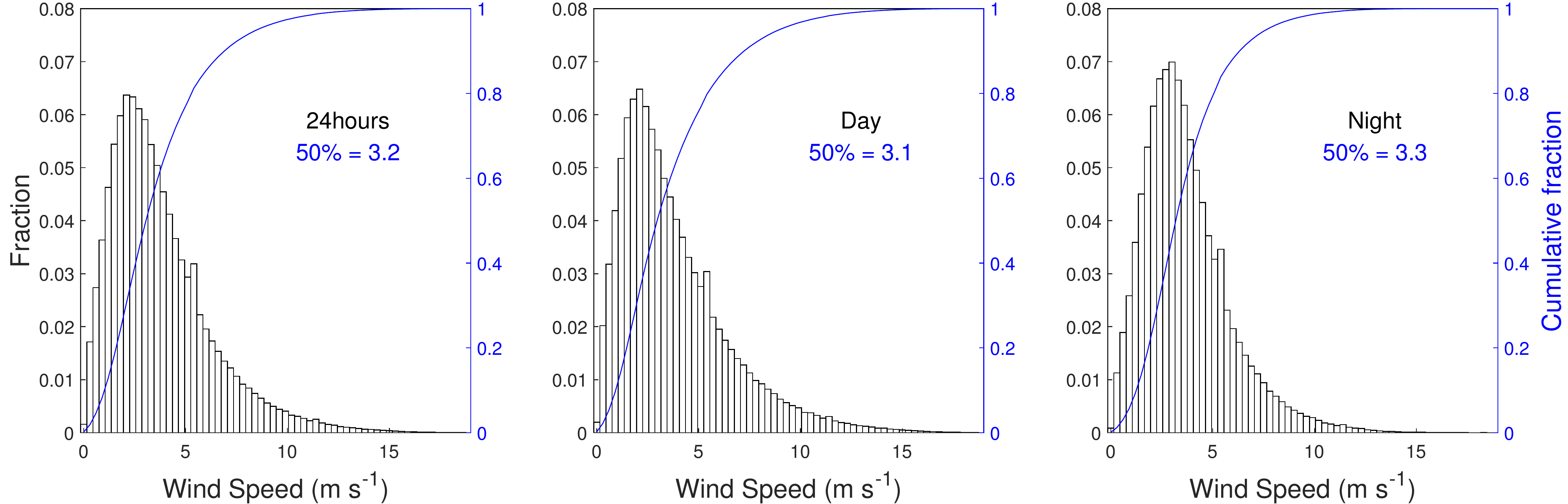}
			\caption{Wind speed  histograms and cumulative distributions on 24 h, daytime and night-time basis from  July 2013 to August 2015.  See the electronic edition of the PASP for a color version of this figure.}
			\label{fig-windSpeed}
		\end{center}
	\end{figure}

\begin{figure}
		\begin{center}
			\includegraphics[angle=0, width=1\textwidth]{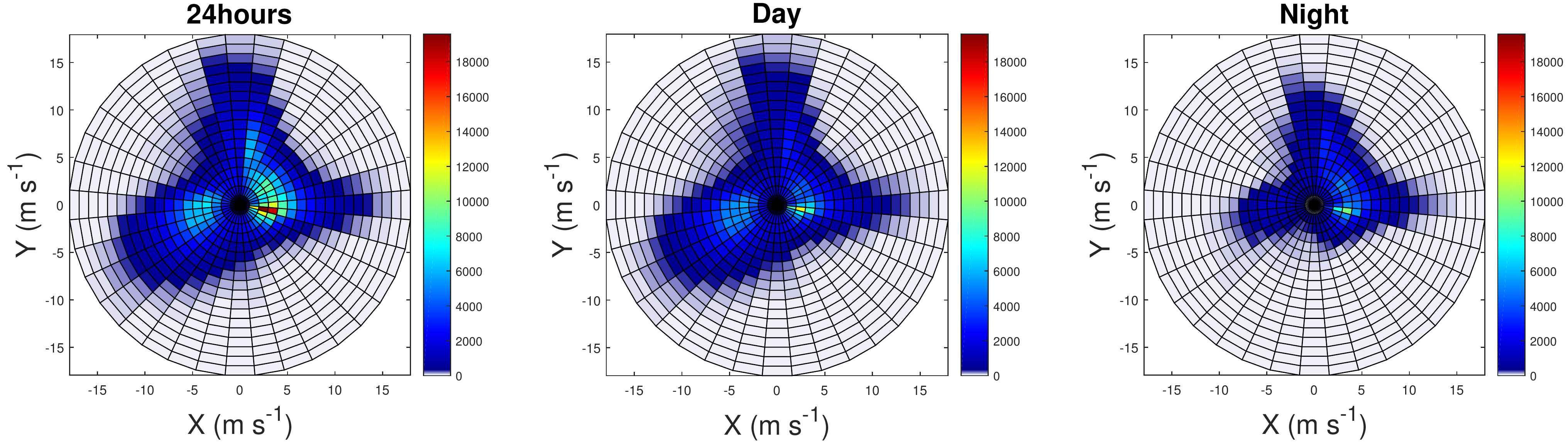}
			\caption{Wind rose density on 24 h, daytime and night-time basis from  July 2013 to August 2015.  North is up and east is right. The color  represents  the number of data points per cell. See the electronic edition of the PASP for a color version of this figure.}
			\label{fig-windDirection}
		\end{center}
	\end{figure}

Wind direction data is available only from the new weather station which was installed in July 2013. Therefore the site statistics regarding wind direction are not as long-term as for the other weather parameters. The median wind velocity at 10 meters above the ground is $3.2~{\rm m~s^{-1}}$ and the typical wind direction is east. We have analyzed the wind speed and direction over daytime, night-time, and 24 h periods. Our definition of night-time is again between astronomical twilights. The distributions and cumulative statistics for the wind speed during these 24 h, daytime and night-time periods are shown in Figure \ref{fig-windSpeed}. We find that the fraction of time during which the wind speed is above $10~{\rm m~s^{-1}}$ is higher during daytime than during night-time. High wind speeds can cause trouble for observations \citep{1985VA.....28..449M}, but we conclude that the loss of observing time due to high wind speeds is not likely to be common on this site. This can also be clearly seen in the wind rose plots (Fig.\ref{fig-windDirection}). We notice that wind directions are mainly from the east during night-time and from both east and west during daytime. The cloud motion in this region is mainly from west to east, but the prevailing wind direction at the ground level is from the east, especially during the night. So perhaps the local ground level wind direction is influenced by local topography. 

Monthly statistics (for the full 24 hour measurements) of wind speed and direction are shown in Figure \ref{fig-windSpeed_month} and Figure \ref{fig-windDirection_month}. The peak of the distribution is around $2.5~{\rm m~s^{-1}}$ and the wind speeds are relatively stronger in May and weaker in January. We find that the wind speed is normally lower than $10~{\rm m~s^{-1}}$ and seldom reaches $15~{\rm m~s^{-1}}$.  However, our recorded maximum wind speed is $18.8~{\rm m~s^{-1}}$, which is an important point to note regarding further precautions and improvement of the facility.  Wind directions are mainly concentrated from the east all the year, and almost never from the south. We don't find other obvious patterns regarding the wind direction.  Notice that the median wind speed is almost constant throughout the year, ranging from $2.7~{\rm m~s^{-1}}$ to $3.8~{\rm m~s^{-1}}$.

	\begin{landscape}
	\begin{figure}
		\begin{center}
			\includegraphics[angle=0, scale= 0.5]{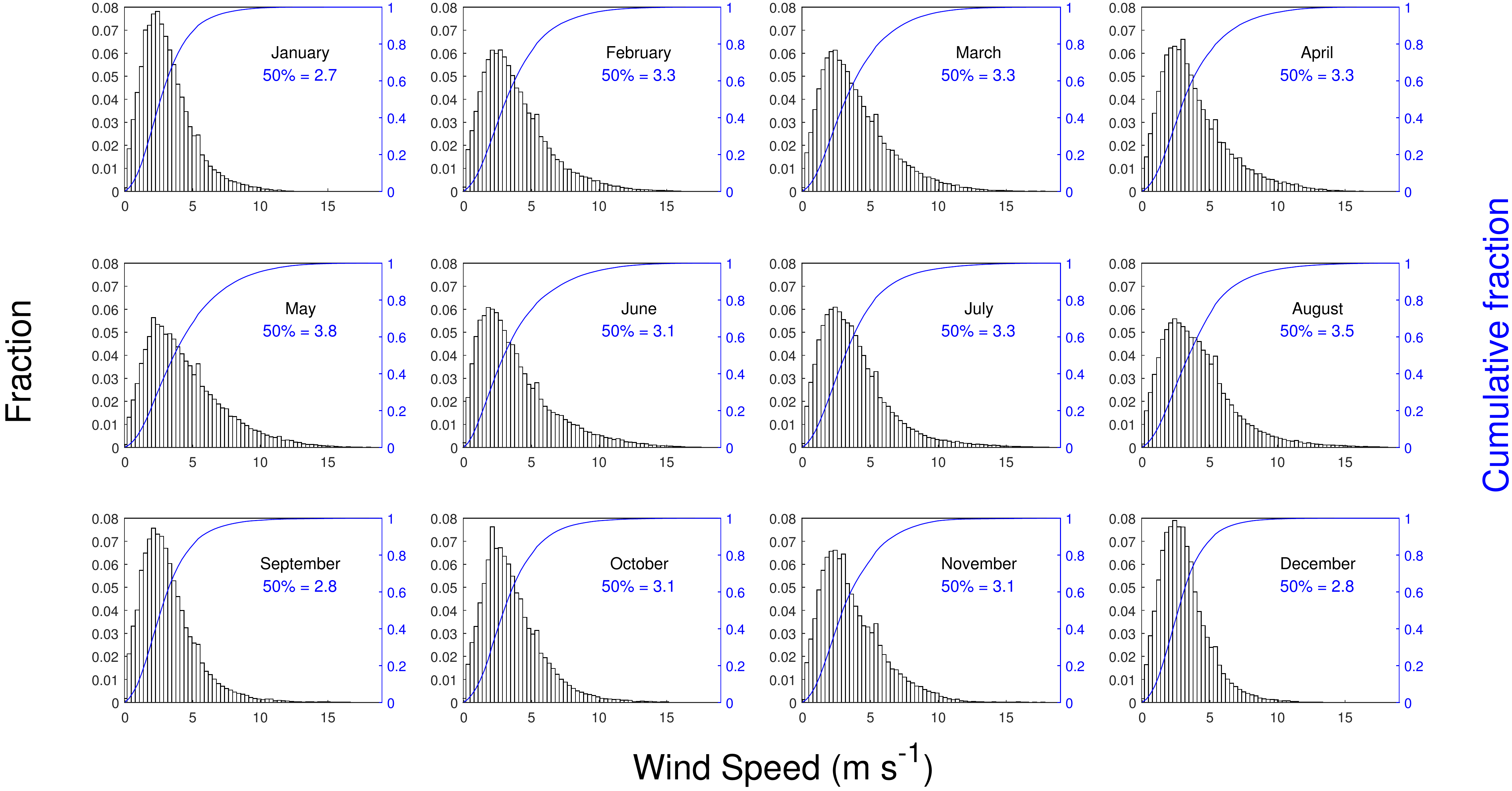}
			\caption{Wind speed  histograms and cumulative distributions by month from  July 2013 to August 2015.  See the electronic edition of the PASP for a color version of this figure.}
			\label{fig-windSpeed_month}
		\end{center}
	\end{figure}
	\begin{figure}
		\begin{center}
			\includegraphics[angle=0, scale= 0.5]{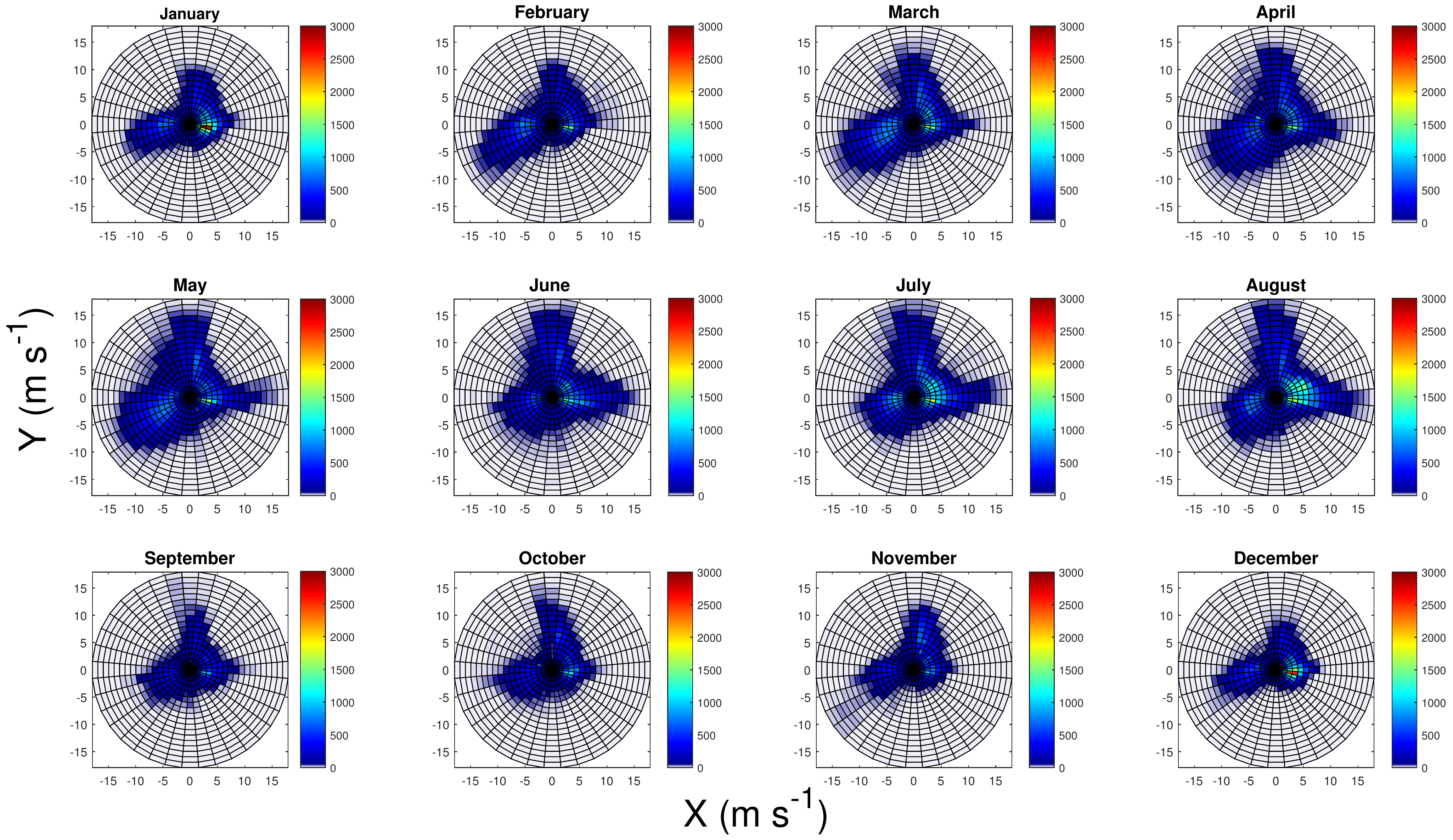}
			\caption{Wind rose density by month from July 2013 to August 2015.  North is up and east is right. The color  represents  the number of data points per cell. See the electronic edition of the PASP for a color version of this figure.}
			\label{fig-windDirection_month}
		\end{center}
	\end{figure}
	\end{landscape}

%\subsection{Statistics on annual observing nights}
\subsection{Statistics on Annual Observing Nights}

 At the beginning of the site survey, an ASC from SBIG, AllSky-340\footnote{\url{http://www.sbig.com/products/cameras/specialty/the-allsky-340-camera/}}, was deployed on the roof for monitoring the night sky and measuring night-time cloud coverage \citep{campbell2010widefield, 2012ASInC...7..187M}. The sensor of ASC has a  300nm (FWHM) bandwidth centered at 480nm. This camera delivers monochrome pictures of the whole sky with a fisheye lens (1.4mm focal length). The image resolution is only $640\times480$ but has a wide view down to the horizon. It has an RS-232 link to a PC for control and data link. The images were downloaded to a local server every five minutes, which were then visually inspected to produce an assessment of the number of clear nights.
This ASC had suffered continued technical problems, usually due to static charges that make it unable to take any images, and has sometimes even been inaccessible. From the summer of 2013, a new system for cloud coverage measurement based on a Canon photo camera was introduced.  This new ASC is composed of a Canon 600D digital SLR camera and a Sigma 4.5 mm f/2.8 fisheye lens. We managed to change the control platform from a Windows$^{\circledR}$  PC to an ARM-based computer running Linux, just like the HRCAM at Dome A \citep{2013IAUS..288...38S}. But our system can save images into FITS files directly to this ARM-based computer without further conversion. Usually conversions such as with cr2fits \citep{cr2fits} involve too heavy computing tasks for ARM systems, so we developed a conversion program in the C language. Exposures are taken every 20 minutes during the daytime and 5 minutes during the night-time and vary from 1/3200 of a second to 30 seconds depending on the night sky brightness\citep{mypapertest}.

Images collected during the night time were used to classify each night into one of three categories:  ``clear'', meaning that it is clear for, at least, 6 hours  \citep{1973PASP...85..255M},   ``cloudy'', meaning 65\% or less cloud coverage, and so the night can be used for spectroscopic observations \citep{1992RMxAA..24..179T}, and useless nights.  Because the first all sky camera did not work continuously, the data volume of the images was incomplete. Data from the observation log, analysis of the SQM data \citep{ClearSkySQM}, and from the weather bureau of Delingha state are used if there are no images during a particular day.  We further require that the wind speed be less than $15~{\rm m~s^{-1}}$ \citep{1985VA.....28..449M} and the humidity be less than $90\%$ \citep{1985VA.....28..449M} to classify a night as either clear or partially cloudy.  Given that the information is judged mainly by visual inspection and meteorological parameters, our results have some uncertainties. The distribution of clear nights by month from 2011 to 2014 is shown in Table \ref{tbl-night}. We find the annual number of clear nights to be about 127.3 and of partial cloudy nights about 120.8. So the combined annual fraction of nights which are useful for observations is found to be  $68.0\%$.

% 2011-04-01   2013-03-07

	\begin{deluxetable}{lccccccccccccccc}
		\tablecolumns{16}
		\tabletypesize{\scriptsize}
%		\rotate
%		\tablewidth{50pc}
		\tablecaption{Statistics on clear skies during night-time at the Delingha site from 2011 to 2014 \label{tbl-night}}
		\tablehead
		{
				\colhead{Year}  &   \colhead{} &  \multicolumn{2}{c}{2011}  &   \colhead{}  & \multicolumn{2}{c}{2012}   &   \colhead{}   &  \multicolumn{2}{c}{2013} &   \colhead{}  & \multicolumn{2}{c}{2014} &   \colhead{}  & \multicolumn{2}{c}{Average}\\
				\cline{1-1} \cline{3-4} \cline{6-7} \cline{9-10} \cline{12-13} \cline{15-16}\\
				\colhead{Month} & \colhead{} & \colhead{Clear}   & \colhead{Cloudy}    & \colhead{} & \colhead{Clear}   & \colhead{Cloudy} & \colhead{} & \colhead{Clear}   & \colhead{Cloudy}& \colhead{}  & \colhead{Clear}   & \colhead{Cloudy} & \colhead{} & \colhead{Clear}   & \colhead{Cloudy}
		 }
		\startdata

			 Jan. & & 10 & 9 &  & 12 & 12 & & 16 & 11 &  & 22 & 3 &  & 15.0 & 8.8\\
			 Feb. & & 12 & 9 &  & 5 & 16 & & 14 & 8 &  & 8 & 10 &  & 9.8 & 10.8\\
			 Mar. & & 10 & 8 &  & 12 & 11 & & 17 & 8 &  & 9 & 14 &  & 12.0 & 10.3\\
			 Apr. & & 17 & 8 &  & 10 & 10 & & 11 & 9 &  & 4 & 14 &  & 10.5 & 10.3 \\
			 May. & & 9 & 9 &  & 7 & 6 & & 7 & 9 &  & 3 & 21 &  & 6.5 & 11.3\\
			 Jun. & & 8 & 8 &  & 4 & 8 & & 6 & 12 &  & 7 & 11 &  & 6.3 & 9.8\\
			 Jul. & & 10 & 7 &  & 4 & 5 & & 4 & 10 &  & 12 & 8 &  & 7.5 & 7.5\\
			 Aug. & & 13 & 5 &  & 5 & 8 & & 11 & 8 &  & 5 & 14 &  & 8.5 & 8.8\\
			 Sep. & & 6 & 16 &  & 12 & 13 & & 9 & 9 & & 8  & 11 &  & 8.8 & 12.3\\
			 Oct. & & 8 & 19 &  & 10 & 18 & & 18 & 7 &  & 14 & 9  &  & 12.5 & 13.3\\
			 Nov. & & 10 & 12 &  & 10 & 8 & & 17 & 8 &  & 14 & 10 &  & 12.8 & 9.5\\
			 Dec. & & 14 & 15 &  & 14 & 6 & & 20 & 7 &  & 21 & 6 &  & 17.3 & 8.5\\
			 total & & 127 & 125 &  & 105 & 121 & & 150 & 106 &  & 127 &131 &  & 127.3 & 120.8\\
%			 \cutinhead{This is a cut-in head}

		\enddata

%		\tablenotetext{a}{with the deluxetable environment}
		\tablecomments{The data from 2011 to the summer of 2013 come from the former ASC and the data from the summer of 2013 to 2014 come from the new ASC}
		
	\end{deluxetable}

%\subsection{The night sky brightness}
\subsection{The Night Sky Brightness}
The sky brightness of the site has been monitored all the time by a commercial device SQM-LE\footnote{\url{http://www.unihedron.com/projects/sqm-le/}} from November 2010. The SQM-LE is low cost and pocket sized, was developed by the Unihedron Company and can measure the sky brightness, providing measures of magnitudes per square arcsecond.  It has an Ethernet interface for convenient data connection. The internal infrared-blocking filter restricts the measurements to the visual bandpass and the FWHM of the angular sensitivity is about $20\,^{\circ}$.  A detailed description of SQM can be found in \citet{cinzano2011night}. This compact device has been used by amateurs and professionals; for instance, \citet{2008ASPC..400..152S} and \citet{2012ASInC...7..187M}. In Delingha, the SQM was put on the tower where the meteorological station was located and pointed to the zenith. At first we used the official Perl application for reading data. However, sometimes the data failed to be recorded until we designed a new program in July 2013 (see section 4 for details). SQM data are sampled every minute.

\begin{figure}
		\begin{center}
			\includegraphics[angle=0, width=1\textwidth]{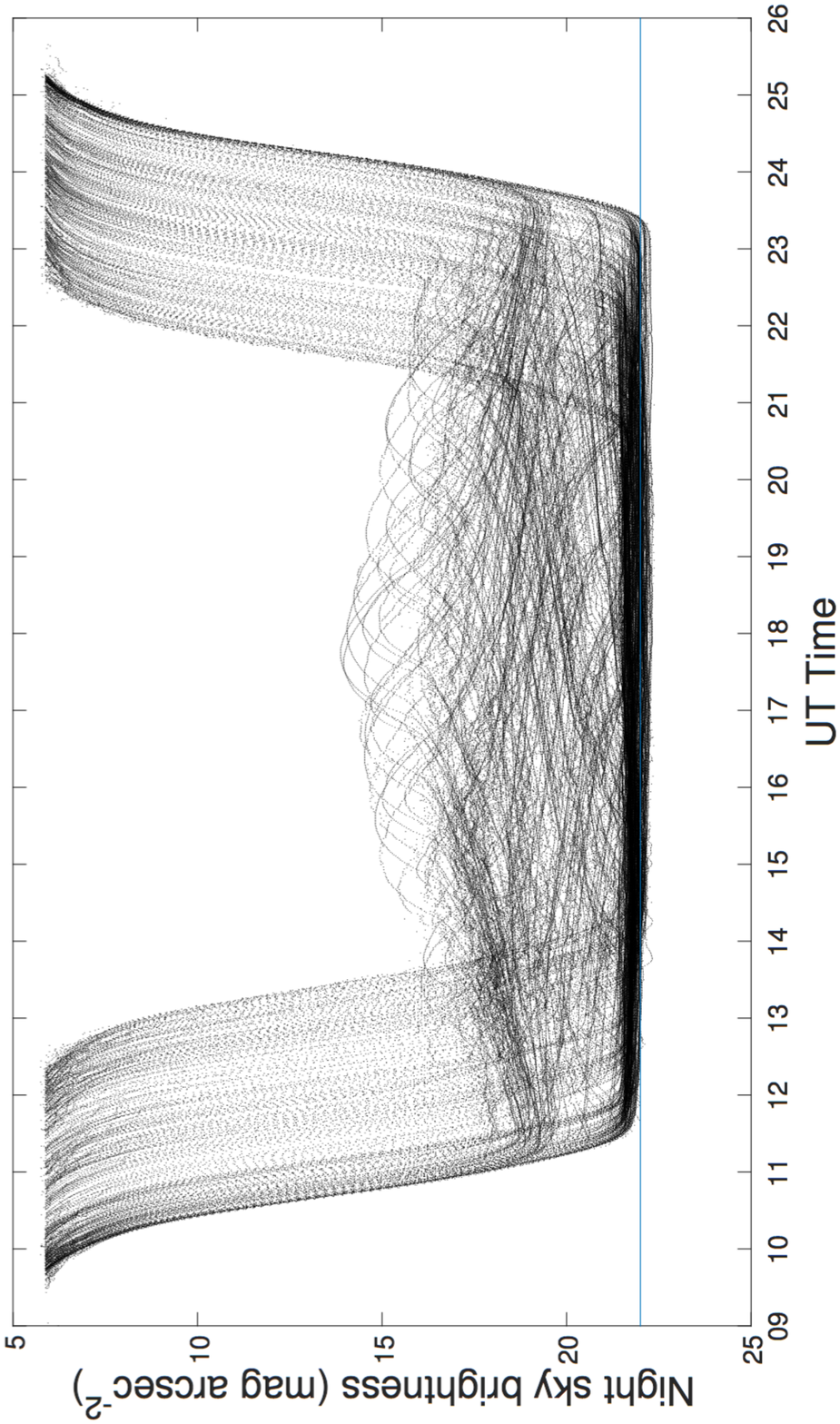}
			\caption{How the sky brightness as measured by our SQM varies as a function of  UT time from  November 2011 to December 2014}
			\label{fig-sqm}
		\end{center}
	\end{figure}

\begin{figure}
		\begin{center}
			\includegraphics[angle=0, width=1\textwidth]{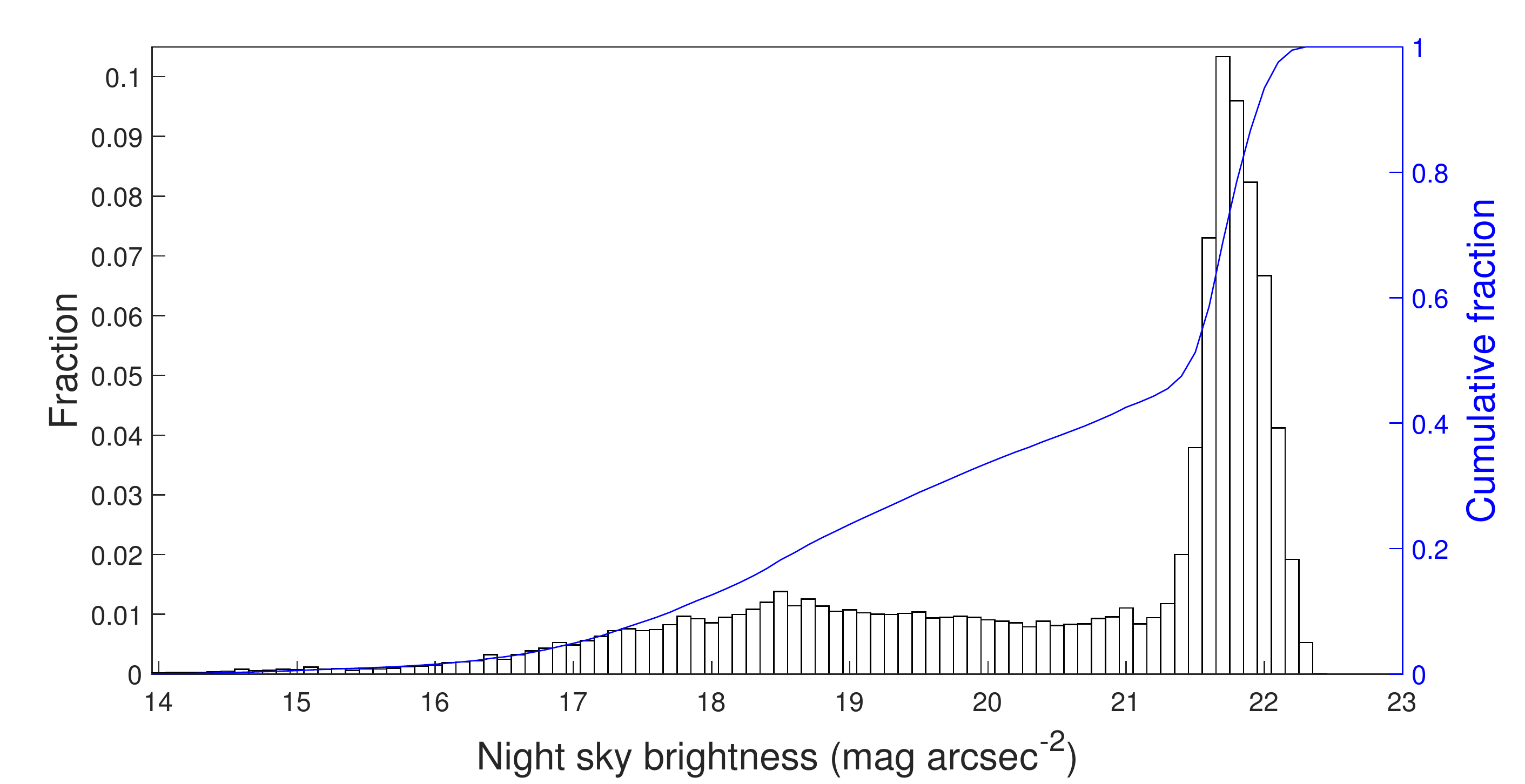}
			\caption{Histograms and cumulative distributions of night sky brightness as measured by our SQM from  November 2011 to December 2014.  See the electronic edition of the PASP for a color version of this figure.}
			\label{fig-sqm_table}
		\end{center}
	\end{figure}

The sky brightness from  November 2010 to December 2014 is plotted in Figure \ref{fig-sqm}. Only the data from clear and partially cloudy nights are used.  It shows that the brightness of the night sky can reach 22 $mag \  arcsec^{-2}$ or fainter in dark phases of the lunar cycle. The influence of moonlight can be clearly seen from this figure. The histograms in Figure \ref{fig-sqm_table} show the night sky brightness distributions and cumulative statistics over the 4 years of our measurements. The values of sky brightness are concentrated between 21.4 and 22.2 $mag \  arcsec^{-2}$ with a median value of  21.5 $mag \  arcsec^{-2}$.

%\subsection{Seeing statistics from 3 year monitor data}
\subsection{Seeing Statistics From 3 Year Monitor Data}
The night-sky seeing was measured by  a Differential Image Motion Monitor (DIMM)  from   2010 to  2012.  The DIMM we used for all measurements was ``home made'', and was placed at the top of the tower where the meteorological station was located. The DIMM made use of a Meade LX200GPS 25 cm telescope,  with the  focal ratio of F/10,  equipped with a Lumenera  SKYnyx2-0   $7.4\ \micron$  square pixels  $480\times 640$ pixel Sony ICX424 CCD camera. Such a configuration gives a pixel scale of 0.61 arcsec. There are two 50 mm apertures on the mask, one with prisms separated by $200$ mm from the other sub-aperture.  The control software and data reduction process are made to run on Windows$^{\circledR}$ XP. The data processing program is developed in C++ and the TCL language with a graphical user interface. Bright stars are the optimal targets for the DIMM.  The CCD exposure time was set to be 10 ms and one seeing value was calculated based on a set of 100 images. The central wavelength of light measured by the DIMM is $ \lambda = 0.5~\rm \mu m$ and  the final results have been corrected for the time zenith angle according to the following relation:

\begin{equation} \label{eq:1}
%	\xout{r_0 = r_{0z}\cdot \left( cosz\right)^{-\frac{3}{5}}  \quad}
	\varepsilon_0 = \varepsilon_{0z}\cdot \left( cosz\right)^{\frac{3}{5}}
\end{equation}
where $z$ is the zenith angle, $\varepsilon_0$ is the seeing at the zenith and  $\varepsilon_{0z}$ is the seeing as determined by our DIMM.

\begin{figure}
	\begin{center}
		\includegraphics[angle=0, width=1\textwidth]{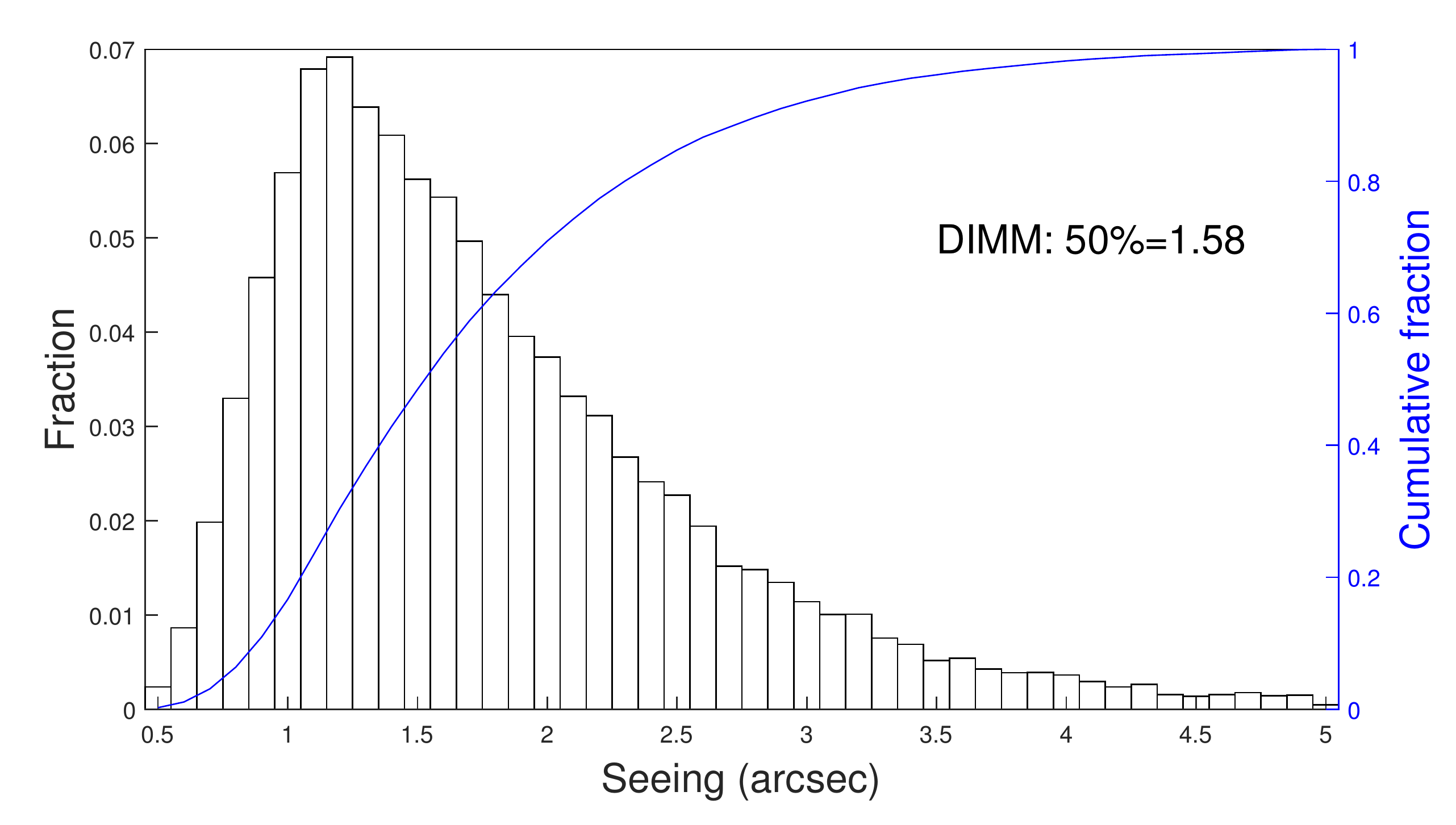}
		\caption{Frequency distribution of DIMM seeing from  2010 to 2012.  See the electronic edition of the PASP for a color version of this figure.}
		\label{fig-dimm_sbig}
	\end{center}
\end{figure}

\begin{landscape}
	\begin{figure}
		\begin{center}
			%		\rotate
			\includegraphics[angle=0,scale= 0.5]{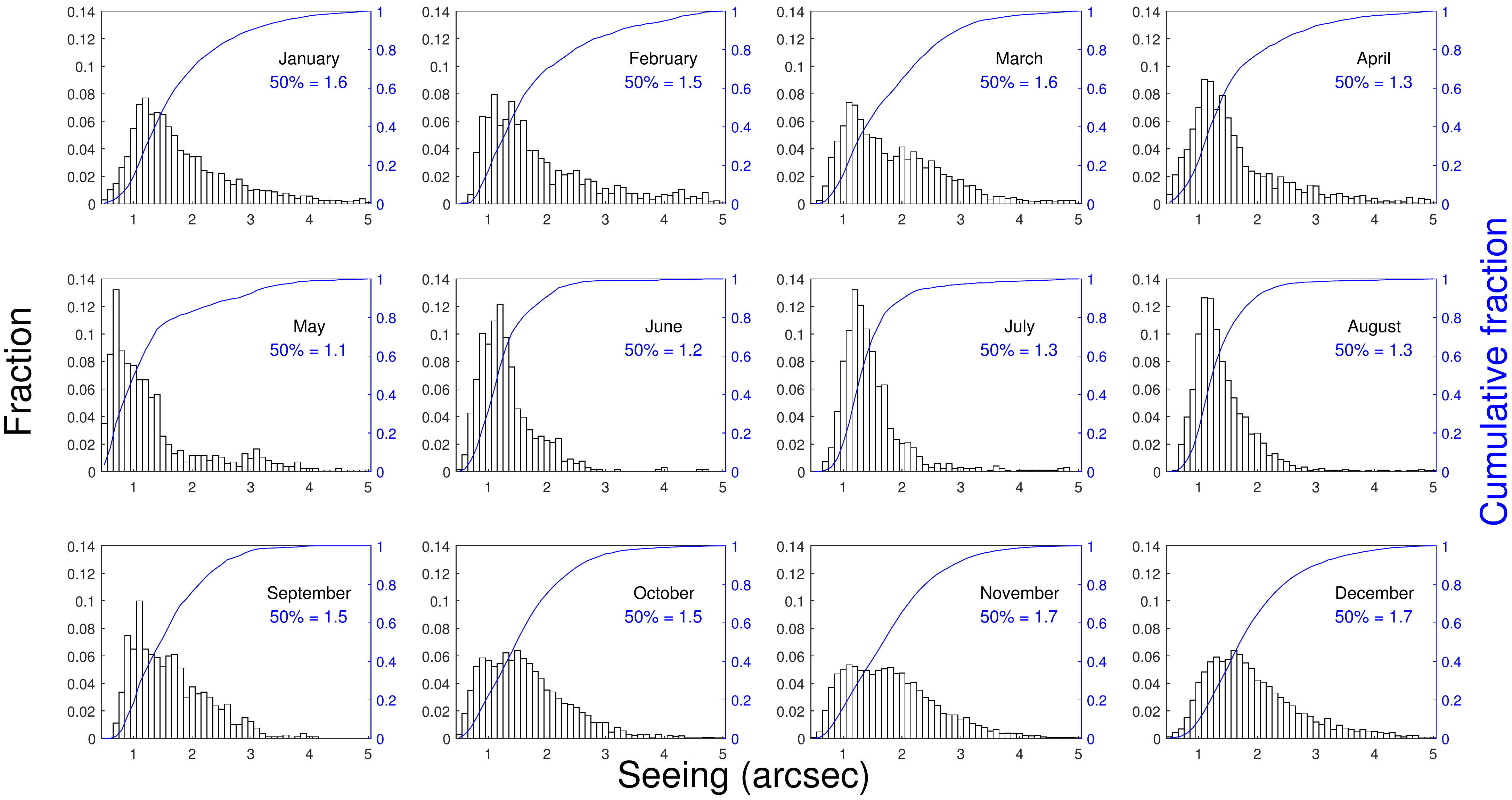}
			%		\scalebox{0.5}{\includegraphics[angle=0]{dimm_sbig1.eps}}
			%			\includegraphics*[0.5cm,1.5cm][0.5cm,1.5cm]{dimm_sbig1.eps}
			%			\includegraphics*[0.5in,8.5in][2.5in,10.5in]{dimm_sbig1.eps}
			\caption{Distributions and cumulative statistics for DIMM seeing by month from 2010 to 2012.  See the electronic edition of the PASP for a color version of this figure.}
			\label{fig-dimm_table}
		\end{center}
	\end{figure}
\end{landscape}

\begin{landscape}
	\begin{figure}
		\begin{center}
			%		\rotate
			\includegraphics[angle=0,scale= 0.50]{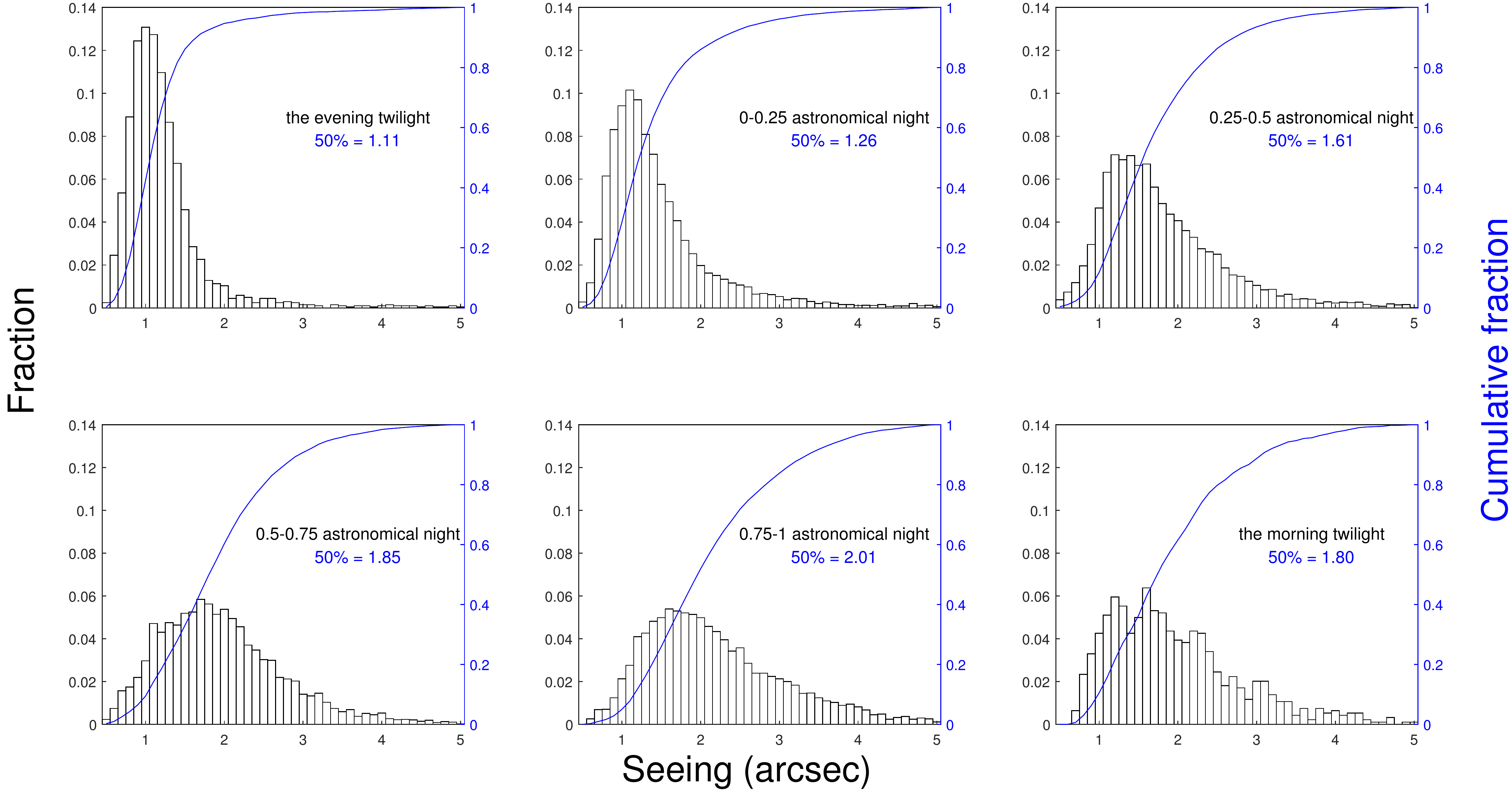}
			%		\scalebox{0.5}{\includegraphics[angle=0]{dimm_sbig1.eps}}
			%			\includegraphics*[0.5cm,1.5cm][0.5cm,1.5cm]{dimm_sbig1.eps}
			%			\includegraphics*[0.5in,8.5in][2.5in,10.5in]{dimm_sbig1.eps}
			\caption{Histograms and cumulative distributions for DIMM seeing  as a function of the time in a day. See the electronic edition of the PASP for a color version of this figure.}
			\label{fig-dimm_sbig1}
		\end{center}
	\end{figure}
\end{landscape}

The distributions and cumulative statistics for DIMM seeing  are presented in Figure \ref{fig-dimm_sbig}, which shows that the median seeing of the DIMM is $1.58~\rm arcsec$, with a distribution peaks around $1.2~\rm arcsec$. The distributions and cumulative statistics for DIMM seeing by month are shown in Figure \ref{fig-dimm_table}. We notice that the seeing distribution has a seasonal dependence, with the median DIMM value better in summer than winter \citep[similar to Xinglong Observatory][]{2012RAA....12..772Y, 2015PASP..127.1292Z}.  Figure \ref{fig-dimm_sbig1} illustrates all data as a function of the time during a day to see intraday variations. Because the length of the astronomical night is different for different times in the year, the astronomical night is split to 4 fractions and each bin, whose length is different along the year, is a quarter of the length of the relevant night. So the whole night is divided into 6 blocks: the evening twilight, 4 fractions of the astronomical night and the morning twilight. The median values of the DIMM measurements show that the seeing become progressively worse during the night (except for the morning twilight).

At the beginning of the site survey, we applied a SBIG seeing monitor\footnote{\url{https://www.sbig.com/products/cameras/specialty/seeing-monitor/}}. It was put on the roof of our lab building, which is 3 m above ground level. That SBIG seeing monitor used a 2.8 cm scope with a focal ratio of F/5.3, equipped with an SBIG ST-402ME CCD camera. The whole device was packed inside a weatherproof container with a clear window on the light path. The lens and box were permanently pointed towards Polaris.  The field of view was large enough that the entire orbit of Polaris about the north celestial pole could be captured. The exposure time of the CCD was set to 5 ms and the seeing values at the zenith were calculated automatically according to the Polaris trail \citep{1965PASP...77..246H}. Most of the seeing values from this SBIG seeing monitor are distributed in the range of from 0.5 to 4.5 arcsec and the median seeing is 2.20  arcsec, peaked around 1.3  arcsec. It can be easily noticed the DIMM seeing values are lower than the SBIG seeing.  This is because the SBIG seeing monitor uses a star trail method which can be easily affected by equipment jitter \citep{2010RAA....10.1061L}.

 We also compared our DIMM setup with other site testing campaigns. There are two DIMMs in regular operation at Xinglong Observatory. One of them uses an 20 cm telescope.  The separation of the sub-aperture in that case is 15 cm, and the exposure time is set to 0.5 ms \citep{2015MNRAS.451.3299L}. All other components are the same as ours. The other DIMM in Xinglong Observatory is slightly different, and uses a 28 cm telescope which is equipped with a AVT Gruppy F-033 CCD camera; the separation of the sub-aperture is 23 cm \citep{2015PASP..127.1292Z}. MASS/DIMM instruments, which are described in detail in \citet{2007MNRAS.382.1268K}, are used for both TMT and E-ELT site testing. For TMT, the MASS/DIMM units are installed on 35 cm telescopes with an SBIG ST7 camera \citep{2009PASP..121.1151S}. Whereas for E-ELT, the MASS/DIMM units are installed on 28 cm telescopes with PCO PixelFly VGA camera \citep{2012PASP..124..868V}.

 Despite those differences in instrument specification, seeing measurements quantitatively reveal site quality. For example, the median seeing of Ventarrones (the E-ELT site)  is 0.91 arcsec  \citep{2012PASP..124..868V}  and Mauna Kea 13N (the TMT site) is  0.75 arcsec  \citep{2009PASP..121.1151S}. Usually sub-arcsec seeing is required for modern optical observing facilities as in almost all the top quality sites. Obviously, the median seeing of 1.58 for the Delingha site cannot compete with the world's best sites. However, considering the science goals and all other limiting conditions for relatively small projects, the seeing statistics we find for Delingha are respectable.

 %For example, the median seeing of Xing-long Observatory is  $ 1.8 \ arcsec$ \citep{2015PASP..127.1292Z}. }

\clearpage

%height=1\textheight   , width=1\textwidth ,scale= 0.55

\clearpage

%\section{Infrastructure and observation management} \label{infra}
%\section{Site monitor system for operations} \label{infra}
\section{SITE MONITOR SYSTEM FOR OPERATIONS} \label{infra}
%what is needed for automation, what have been done, what is to be done. how it does work, .... etc.

In order to help the isolated SONG-China node to run in an automated mode,  instruments for detecting site parameters (cloud coverage, meteorological data, seeing, dust, atmospheric extinction etc.)  should be installed. In addition to those site parameters for observations, more temperature and mechanical sensors for detecting the state of the dome, telescope and spectrograph are needed. Real-time video surveillance is also required for remote control and maintenance. Those additional devices and sensors combine to make an independent monitor system that will be integrated in and serve the operations of the stand-alone node.

\subsection{The specifications of the monitor system}

These instruments and sensors are produced by different companies with different input and output interfaces. Many of them have to be customized before they can be integrated into automated applications. These off-the-shelf items are usually Windows$^{\circledR}$ based, and suffer data instabilities due to communication, software and even power problems, just as shown in Section \ref{condition}. In most cases, the control software of the instruments  writes data of various format into text files. Such information has to be re-handled before being fed to the automated operation of SONG and 50BiN. The goals and tasks are specified as follows:

%		\flushleft
			\begin{enumerate}[(1)]
			
					\item Unified input and output interface
					
					There are four different types of interfaces provided by different instruments current in use at Delingha, namely RS232  and RS485 serial, USB and  Ethernet ports. For the convenience of data connection, we exclusively use an Ethernet interface. Therefore all the instruments and sensors have been adapted to transmit information through the network based on a TCP/IP protocol.
					
					\item Integrity and accuracy of data taking
					
					In order to avoid abnormal termination of the data collection, both the software and hardware of the instruments have been improved.

					\item Data storage
					
					In order to facilitate efficient data management, all the data are pipelined into a database rather than a collection of text files.
					In such a way, the site quality data can be easily queried over the Internet from anywhere. A real-time backup onto another computer is also made, just as for the main scientific data.

					\item Web interface for data display
					
					A visualization of the data is made available through a web-page, which supports interactive data access in real time.  This will help astronomers to get the site information whenever necessary.

		\end{enumerate}

\subsection{Implementation and  applications}

There are many different ways to convert other ports to Ethernet port, for example an RS232 to Ethernet conversion module. We decided to use single board computers with embedded technology. Using such embedded technology has the following advantages: First, many single board computers have the four types of interface at the same time. Second, the control program can be put into the single board computer (which is small sized, low power, and cheap), enabling the board to organise the collection of data from the instruments instead of using a big desktop PC.   Third, because such embedded systems are designed for specific tasks, they usually have high stability compared to a desktop PC.   In addition, this configuration reduces the communication distance between the instrument and the controlling computer; a long range serial or USB connection tends to make communication very unstable.  But the disadvantage is that the control program should be rewritten to make it run on the embedded system.

We take the following measures to ensure data integrity. Firstly, the power supply for the devices is through a small 12V uninterruptible power supply (UPS). When power supply from the grid fails, the instruments can continue to work through this small UPS. Secondly, the control program running in the single board computer is divided into two parts: a daemon and a worker/slave process. The daemon is a background process which boots up automatically when the single board computer is powered on. The daemon process can call the worker process, which is in charge of controlling the instrument.  When called by the daemon, the worker process will initialize the instrument, complete the data acquisition, check the date, send the data to the database and then exit. If the data cannot be sent to the database, the data will be saved in the local flash memory of the single board computer. This situation occurs rather frequently due to network instability and/or hardware problems.  When the network is restored, the data will then be sent to the database.

We use two database servers to mirror the data of the monitoring system. One is at  the Delingha site and the other is in Beijing. If one is broken, data can be recovered from another. We use FusionCharts\footnote{\url{http://www.fusioncharts.com/}} to clearly visualize the data and use Asynchronous JavaScript and XML (AJAX) to support the user's interaction with the data sources.

\begin{figure}
	\begin{center}
		\includegraphics[angle=0, width=1\textwidth]{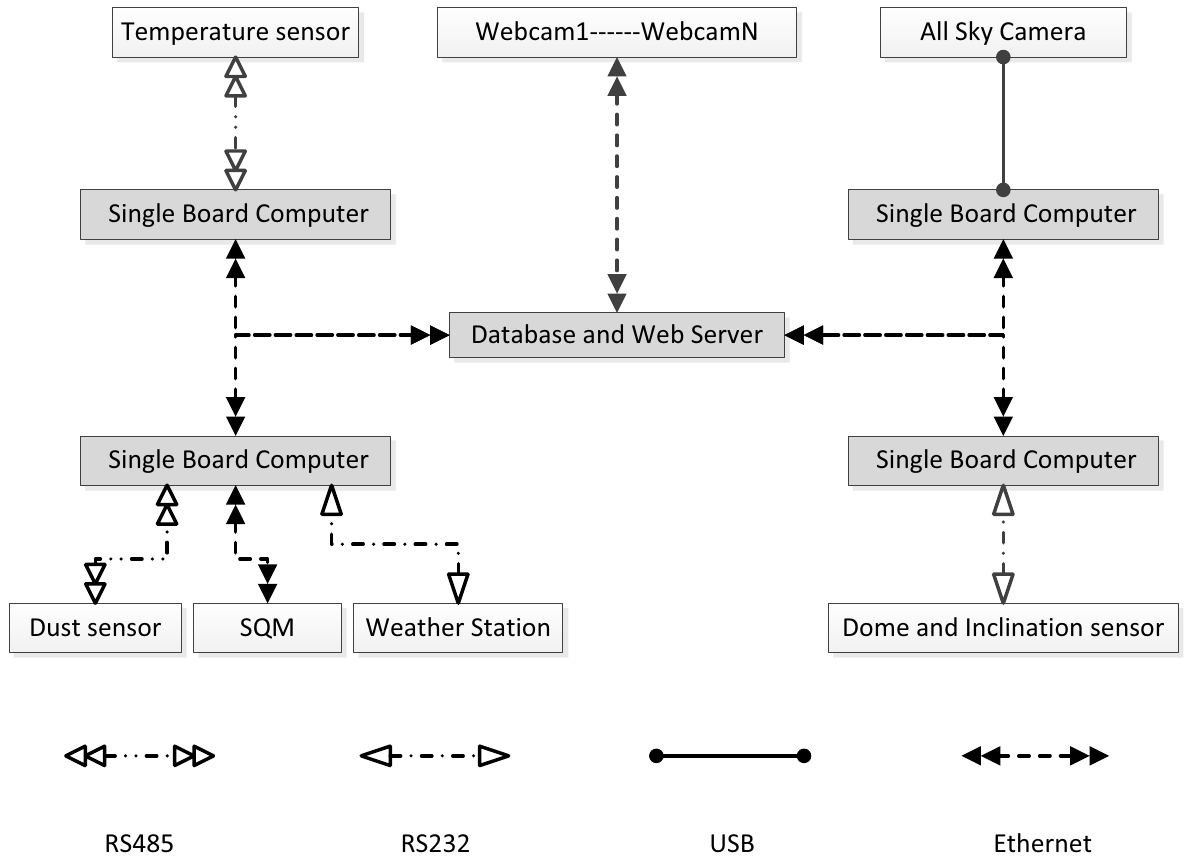}
		\caption{ Schematic layout of the monitor system. Connection types are coded using the symbols at the bottom of the schematic.}
		\label{fig-hardware}
	\end{center}
\end{figure}

Figure \ref{fig-hardware} is a schematic layout of the interconnections required for the current monitor system. We have finished most of the development design for the instruments and sensors of the monitor system using embedded technology. Those completed instruments and sensors can work in an automated mode and can transmit data through the network. A unified program module and class were defined in this process which was designed such that new instruments should be able to be easily added to the monitor system.

%\section{Summary and Conclusions} \label{conclusion}
\section{SUMMARY AND CONCLUSIONS} \label{conclusion}

Based on long time surveillance of the critical site parameters at the Delingha site, the SONG international alliance have agree to the final selection of Delingha for the location of the Chinese node. The site qualification work presented in the paper can be summarized in the following:

\begin{enumerate}

\item We have visually analyzed 155171 frames of ASC images collected over 4 years, on top of the weather data from local meteorological records. The site has an annual average number of 127.3 photometric nights and 120.8 non-photometric but observable nights.  So an average of $68.0\%$ of the nights can be used for observing with $34.9\%$ classified as clear nights.

\item Our weather stations collected over 1.8 million data points continuously through 2010--2014. These indicate that the annual average temperature is $1.8\,^{\circ}\mathrm{C}$ with a maximum of $29.4\,^{\circ}\mathrm{C}$ and minimum of $-31.0\,^{\circ}\mathrm{C}$. The average value of the relative humidity is $33.1\%$. The humidity is usually less than $50\%$, with a median value of $32.6\%$.  Although we found that relative humidity during night-time is higher than that during daytime, the relative night-time humidity still has a median value of only $37.1\%$.  This low relative humidity is favourable for astronomical observations. The median wind velocity  is $3.2~{\rm m~s^{-1}}$ and the peak value of the distribution is around $2.5~{\rm m~s^{-1}}$.  Most of the time the wind speed is lower than $10~{\rm m~s^{-1}}$ and seldom reaches $15~{\rm m~s^{-1}}$. The fraction of the time when the wind speed is above  $10~{\rm m~s^{-1}}$ is higher during daytime than than during night-time.  The wind speed distribution varies very little from month to month, with median speeds ranging from $2.7~{\rm m~s^{-1}}$ to $3.8~{\rm m~s^{-1}}$, which are good values for astronomical observations.

\item The brightness of the night sky can reach 22 $mag ~\rm  arcsec^{-2}$ without contamination from moonlight, and the sky brightness measurements are concentrated between 21.4 and 22.2 $mag ~\rm  arcsec^{-2}$ with a median value of  21.5 $mag ~\rm  arcsec^{-2}$.  There is some light pollution near the west horizon due to the fast development of the city Delingha, but the site is overall still very good in terms of night sky brightness.

\item From over 236 thousand data points collected in the period from 2010--2012, mostly continuously observed during clear or partially clear nights, we presented seeing statistics from both SBIG and DIMM instruments. The median seeing determined from DIMM is 1.58  arcsec whereas the distribution peaks around 1.2  arcsec. The median seeing from the SBIG monitor is 2.20  arcsec whilst the distribution peaks around 1.3  arcsec. The median DIMM seeing value is better in summer than winter, and there is a trend in the DIMM data indicating that the seeing is typically better at the beginning of the night than at the end of the night.
%The DIMM data shows that the site has a median value of seeing around 1.5arcsec.

\end{enumerate}

Since these statistics we have presented regarding the key site parameters are based on data spanning several annual cycles, we conclude that the site has been characterized reliably. The question to be answered is whether this site is appropriate for SONG/50BiN science goals. The most obvious shortcoming is the seeing. With a median seing value of about 1.5 arcsec,  a trade-off has to be made in science missions between slit width (therefore spectral resolution) and overall throughput for SONG, while it is between spatial resolution and S/N for 50BiN (photometry). Fortunately, the final physical quantity we want to derive from stellar spectra is radial velocity, which depends more on efficiency (or S/N) than resolution. Test observations have shown that by slightly compromising between the slit width and exposure time, we can deliver radial velocity measurements comparible to the SONG prototype node on Tenerife and therefore produce the same quality of science data. For photometry with 50BiN, the seeing becomes critical only for very crowded fields. For most of the planned targets of 50BiN, this is not a significant issue. We have a pixel scale of 0.57 on the 50BiN detectors, which was actually designed for the specific seeing measured at the site. 
With regard to the advantages of selecting the site for a SONG node, it is firstly the existing site with the best infrastructure in the most appropriate range of longitudes to meet the requirements of the network in the northern hemisphere.  Moreover, for a SONG node the number of clear nights is a much more important factor than the seeing.   Of course Delingha is not competitive with the world's top sites for optical astronomy, but it is at least one of the best places for a SONG node in east-Asia.  Combining all of the limiting conditions, including the project's budget, we can say that this site meets the basic requirements of SONG.  The Delingha site is also typical in terms of the general climate pattern for astronomical observations on the Qinghai-Tibetan Plateau, and can be regarded as a reference for current and future site surveys of this area.

In order to provide real-time surveillance of the site conditions during the operation of both SONG and 50BiN, we have developed a monitoring system. Using embedded technology we have converted the output of the individual instruments to Ethernet interfaces. All the data are acquired by the control program running on those embedded systems, except for video streams. The data are directly saved to a database and can be accessed and visualized from the internet. This monitoring data will help to ensure the quality of science output from these programs, and of course will also serve all future instrumentation that will come to this site.

\section*{Acknowledgements}
This work is support by MOST  973 grant  2014CB845700, the Strategic Priority Research Program ``The Emergence of Cosmological Structures'' of the Chinese Academy of Sciences (Grant No. XDB09000000), and NSFC grants 11473037, 11303021,11373037. The authors wish to thank Delingha site of Purple Mountain Observatory for continuous support since 2009, and all members in the team, especially the night assistants. Thank our colleague, Stephen Justham, for language proof reading the manuscript.

                     %% AASTeX
%\bibliography{reference}
%\bibliography{example1}

\end{document}